\documentclass[10pt,fullpage]{article}

\usepackage{fullpage}
\usepackage{times}
\usepackage{amsmath}
\usepackage{amssymb}
\usepackage{amstext}
\usepackage{epsfig}
\usepackage{graphicx}
\usepackage{placeins}
\usepackage{url}
\usepackage{subfig}
\usepackage{paralist}
\usepackage{caption}
\usepackage{multirow}
\usepackage{algorithm}
\usepackage{algorithmic}

\newcommand{\eat}[1]{}

\makeatletter
\newenvironment{sql}%
 {\vskip 5pt\begin{list}{}{%
  \setlength{\topsep}{0pt}\setlength{\partopsep}{0pt}\setlength{\parskip}{0pt}%
  \setlength{\parsep}{0pt}\setlength{\labelwidth}{0pt}%
  \setlength{\rightmargin}{0pt}\setlength{\leftmargin}{0pt}%
  \setlength{\labelsep}{0pt}%
  \obeylines\@vobeyspaces\normalfont\ttfamily%
  \item[]}}
 {\end{list}\vskip5pt\noindent}
\makeatother

\begin{document}

\date{December 2014}

\title{Speculative Approximations for Terascale Analytics}

\author{
Chengjie Qin \hspace*{2cm} Florin Rusu\\
       \small{University of California, Merced}\\
       \small{5200 N Lake Road}\\
       \small{Merced, CA 95343}\\
       \small\texttt{\{cqin3,frusu\}@ucmerced.edu}
}

\maketitle

\begin{abstract}

\textit{Model calibration} is a major challenge faced by the plethora of statistical analytics packages that are increasingly used in Big Data applications. Identifying the optimal model parameters is a time-consuming process that has to be executed from scratch for every dataset/model combination even by experienced data scientists. We argue that the incapacity to evaluate multiple parameter configurations simultaneously and the lack of support to quickly identify sub-optimal configurations are the principal causes.

In this paper, we develop two database-inspired techniques for efficient model calibration. \textit{Speculative parameter testing} applies advanced parallel multi-query processing methods to evaluate several configurations concurrently. The number of configurations is determined adaptively at runtime, while the configurations themselves are extracted from a distribution that is continuously learned following a Bayesian process. \textit{Online aggregation} is applied to identify sub-optimal configurations early in the processing by incrementally sampling the training dataset and estimating the objective function corresponding to each configuration. We design concurrent online aggregation estimators and define halting conditions to accurately and timely stop the execution.

We apply the proposed techniques to \textit{distributed gradient descent optimization} -- batch and incremental -- for support vector machines and logistic regression models. We implement the resulting solutions in GLADE PF-OLA -- a state-of-the-art Big Data analytics system -- and evaluate their performance over terascale-size synthetic and real datasets. The results confirm that as many as 32 configurations can be evaluated concurrently almost as fast as one, while sub-optimal configurations are detected accurately in as little as a $1/20^{\text{th}}$ fraction of the time.

\end{abstract}

\section{Introduction}\label{sec:intro}

Big Data analytics is a major topic in contemporary data management and machine learning research and practice. Many platforms, e.g., OptiML~\cite{optiml}, GraphLab~\cite{graphlab,graphlab-distributed,graphlab-disk}, SystemML~\cite{systemml}, SimSQL~\cite{simsql}, Google Brain~\cite{google-brain}, GLADE~\cite{glade:osr,glade:sigmod} and libraries, e.g., MADlib~\cite{madlib}, Bismarck~\cite{bismarck}, MLlib~\cite{mllib}, Vowpal Wabbit~\cite{vowpal-wabbit}, have been proposed to provide support for distributed/parallel statistical analytics.

\textit{Model calibration} is a fundamental problem that has to be handled by any Big Data analytics system. Identifying the optimal model parameters is an interactive, human-in-the-loop process that requires many hours -- if not days and months -- even for experienced data scientists. From discussions with skilled data scientists and our own experience, we identified several reasons that make model calibration a difficult problem. The first reason is that the entire process has to be executed from scratch for every dataset/model combination. There is little to nothing that can be reused from past experience when a new model has to be trained on an existing dataset or even when the same model is applied to a new dataset. The second reason is the massive size of the parameter space---both in terms of cardinality and dimensionality. Moreover, the optimal parameter configuration is dependent on the position in the model space. And third, parameter configurations are evaluated iteratively---one at a time. This is problematic because the complete evaluation of a single configuration -- even sub-optimal ones -- can take prohibitively long.

\paragraph*{Motivating example.}
\textit{Gradient descent optimization}~\cite{bertsekas:igd} is a fundamental method for model calibration due to its generality and simplicity. It can be applied virtually to any analytics model~\cite{bismarck} -- including support vector machines (SVM), logistic regression, low-rank matrix factorization, conditional random fields, and deep neural networks -- for which the gradient or sub-gradient can be computed or estimated. All the statistical analytics platforms mentioned previously implement one version or another of gradient descent optimization. Although there is essentially a single parameter, i.e., the step size, that has to be set in a gradient descent method, its impact on model calibration is tremendous. Finding a good-enough step size can be a time-consuming task. More so, in the context of the massive datasets and highly-dimensional models encountered in Big Data applications.

The standard practice of applying gradient descent to model calibration, e.g., MADlib, Vowpal Wabbit, MLlib, Bismarck, illustrates the identified problems perfectly. For a new dataset/model combination, an arbitrary step size is chosen. Model training is executed for a fixed number of iterations. Since the objective function is computed only for the result model -- due to the additional pass over the data it incurs -- it is impossible to identify bad step sizes in a smaller number of iterations. The process is repeated with different step values, chosen based on previous iterations, until a good-enough step size is found. Certain systems, e.g., Google Brain, support the evaluation of multiple step sizes concurrently. This is done by executing independent jobs on a massive cluster, without any sort of sharing.

\paragraph*{Problem statement.}
We consider the abstract problem of \textit{distributed model calibration with iterative optimization methods}, e.g., gradient descent. We argue that the incapacity to evaluate multiple parameter configurations simultaneously and the lack of support to quickly identify sub-optimal configurations are the principal causes that make model calibration difficult. It is important to emphasize that these problems are not specific to a particular model, but rather they are inherent to the optimization method used in training. The target of our methods is to find optimal configurations for the tunable hyper-parameters of the optimization method, e.g., step size, which, in turn, facilitate the discovery of optimal values for the model parameters. Therefore, we investigate \textit{speculative processing} and \textit{intra-iteration approximations} for the optimization of distributed model calibration. Speculative iteration processing allows concurrent evaluation of multiple parameter configurations in a single pass over the training data---without the proportional increase in running time. Intra-iteration approximation allows for faster convergence detection---and corresponding reduction in iteration time. When put together, these techniques have the potential to reduce model calibration time significantly.

\paragraph*{Contributions.}
In this paper, we develop two database-inspired techniques for efficient model calibration. \textit{Speculative parameter testing} applies advanced parallel multi-query processing methods to evaluate several configurations concurrently. \textit{Online aggregation} is applied to identify sub-optimal configurations early in the processing by incrementally sampling the training dataset and estimating the objective function corresponding to each configuration.

Our major contribution is the conceptual integration of parallel multi-query processing and approximation for efficient and effective large-scale model calibration with gradient descent methods. This requires novel technical solutions as well as engineering artifacts in order to bring significant improvements to the state-of-the-art. Specific contributions include:
\begin{compactitem}
\item We design speculative parameter testing algorithms that evaluate multiple configurations simultaneously. The number of configurations is determined adaptively and dynamically at runtime. The configurations are drawn from a parametric distribution that is continuously updated using a Bayesian statistics~\cite{bayes-stat} approach.
\item We design concurrent online aggregation estimators and define halting conditions to accurately and timely stop the execution. We provide efficient parallel solutions for the evaluation of the speculative estimators and of their corresponding confidence bounds that guarantee fast convergence.
\item We apply these techniques to distributed gradient descent optimization for SVM and logistic regression models. As a prerequisite, we formalize gradient descent optimization as a database aggregation problem.
\item We provide an extensive comparison between batch and incremental gradient descent that reveals that -- contrary to the generally accepted opinion -- batch gradient descent methods are better suited for distributed processing over massive datasets due to their linearity properties. With the proposed techniques integrated, batch gradient descent always outperforms the incremental method both in convergence speed and execution time.
\item We implement the proposed solutions in GLADE PF-OLA~\cite{glade:osr,glade:sigmod,pfola:dapd,pfola:demo} -- a state-of-the-art Big Data analytics system -- and evaluate the performance over terascale-size synthetic and real datasets. The results confirm that as many as 32 configurations can be evaluated concurrently almost as fast as one, while sub-optimal configurations can be detected in a $1/20^{\text{th}}$ fraction of the time. These translate in more than 100X faster convergence over MLlib on Spark~\cite{mllib} and 2X over Vowpal Wabbit~\cite{vowpal-wabbit} on a terascale-size real dataset. 
\end{compactitem}

\paragraph*{Outline.}
The model calibration problem is presented in Section~\ref{sec:problem}. Gradient descent solutions are introduced in Section~\ref{sec:grad-descent}, while distributed gradient descent is discussed in Section~\ref{sec:par-grad-descent}. Speculative parameter testing is presented in Section~\ref{sec:multiple-steps}. Intra-iteration approximation with online aggregation is detailed in Section~\ref{sec:ola-grad-descent}. Experimental results that evaluate thoroughly the efficiency and efficacy of the proposed methods and compare the GLADE PF-OLA implementation against the state-of-the-art are presented in Section~\ref{sec:experiments}. Section~\ref{sec:rel-work} discusses relevant related work, while Section~\ref{sec:conclusions} concludes the paper.

\section{Problem Formulation}\label{sec:problem}

Consider the following model calibration problem with a linearly separable objective function:
\begin{equation}\label{eq:optim-form}
\Lambda(\vec{w}) = \textit{min}_{w \in \mathbb{R}^{d}} \sum_{i=1}^{N} f\left(\vec{w}, \vec{x_{i}}; y_{i}\right) + \mu R(\vec{w})
\end{equation}
in which a $d$-dimensional vector $\vec{w}$, $d \geq 1$, known as the model, has to be found such that the objective function is minimized. The constants $\vec{x_{i}}$ and $y_{i}$, $1 \leq i \leq N$, correspond to the feature vector of the $\text{i}^{\text{th}}$ data example and its scalar label, respectively. Function $f$ is known as the loss while $R$ is a regularization term to prevent overfitting. $\mu$ is a constant. For example, the objective function in SVM classification with -1/+1 labels and $\text{L}_{\text{1}}$-norm regularization is given by $\sum_{i}{\left(1-y_{i} \vec{w}^{T}\cdot \vec{x_{i}}\right)_{+}} + \mu ||\vec{w}||_{1}$.

Gradient descent represents, by far, the most popular method to solve the class of optimization problems given in Eq.~(\ref{eq:optim-form}). Gradient descent is an iterative optimization algorithm that starts from an arbitrary point $\vec{w}^{(0)}$ and computes new points $\vec{w}^{(k+1)}$ such that the loss decreases at every step, i.e., $f(w^{(k+1)}) < f(w^{(k)})$. The new points $\vec{w}^{(k+1)}$ are determined by moving along the opposite $\Lambda$ gradient direction. Formally, the $\Lambda$ gradient is a vector consisting of entries given by the partial derivative with respect to each dimension, i.e., $\nabla \Lambda(\vec{w}) = \left[\frac{\partial\Lambda(\vec{w})}{\partial{w_{1}}}, \dots, \frac{\partial\Lambda(\vec{w})}{\partial{w_{d}}}\right]$. Computing the gradient for the formulation in Eq.~(\ref{eq:optim-form}) reduces to the gradient computation for the loss $f$ and the regularizer $R$, respectively. The length of the move at a given iteration is known as the step size, denoted by $\alpha^{(k)}$. With these, we can write the recursive equation characterizing any gradient descent method:
\begin{equation}\label{eq:grad-step}
\vec{w}^{(k+1)} = \vec{w}^{(k)} - \alpha^{(k)} \nabla \Lambda\left(\vec{w}^{(k)}\right)
\end{equation}
In order to check for convergence, the objective function $\Lambda$ has to be evaluated at $\vec{w}^{(k+1)}$ after each iteration. Convergence to the global minimum is guaranteed only when $\Lambda$ is convex. This implies that both the loss $f$ and the regularizer $R$ are convex.

The specific problem we consider in this paper is how to solve the optimization formulation given in Eq.~(\ref{eq:optim-form}) using generic gradient descent methods when the training dataset consisting of $N$ (vector, label) pairs $\{(\vec{x}_{1}, y_{1}), \dots, (\vec{x}_{N}, y_{N})\}$ is partitioned into $M$ subsets. Each subset is assigned to a different processing node for execution. We focus on the case when $N$ is extremely large and each of the $M$ subsets are disk-resident. Gradient and loss computation can take a prohibitive amount of time in this case. Moreover, gradient descent methods are highly sensitive to a series of parameters that require careful tuning, i.e., repeated execution with different configurations, for every dataset. Thus, novel techniques are required in order to scale gradient descent to the largest Big Data models.

\section{Gradient Descent Methods}\label{sec:grad-descent}

In this section, we introduce the basic gradient descent optimization algorithms. We start with the standard batch gradient descent algorithm which is a direct implementation of the theory. Then, we present incremental or stochastic gradient descent, a popular alternative tailored for datasets containing a large number of examples. In addition to these primitive methods, we also discuss two derived algorithms---coordinate descent and L-BFGS. We conclude the section with a comparison of the two standard gradient descent approaches.

\subsection{Batch Gradient Descent}\label{ssec:grad-descent:batch}

The pseudo-code for gradient descent is given in Algorithm~\ref{alg:batch-gd}. This is also known as batch gradient descent (BGD). The algorithm takes as input the data examples and their labels, the loss function $f$ and the gradient of the objective $\nabla\Lambda$, and initial values for the model and step size. The optimal model is returned. The main stages are gradient computation and model and step size update. They are executed until convergence is achieved. Convergence can be specified as a fixed number of iterations or based on the loss, e.g., the loss difference across consecutive iterations decreases below a given threshold. In the later case, the loss has to be computed after every iteration, which incurs an additional pass over the data.

\paragraph*{Model and step size update.}
The standard approach to compute the updated model $\vec{w}^{(k+1)}$, once the direction of the gradient is determined, is to use line search methods~\cite{convex-optimization}. $\vec{w}^{(k+1)}$ is found by iteratively trying different step sizes along the opposite gradient direction until the decrease in loss is above a user-defined threshold, i.e., the Wolfe conditions~\cite{convex-optimization}. Line search achieves a tradeoff between optimality and runtime by choosing only an approximation to the optimal step size, but in shorter time. Nonetheless, line search is still iterative in nature and requires objective and gradient loss evaluation. These involve multiple passes over the entire data. A widely used alternative is to fix the step size to some arbitrary value and then decrease it as more iterations are executed, i.e., $\alpha^{(k)} \rightarrow 0$ as $k \rightarrow \infty$. The initial step size $\alpha^{(0)}$ and the decay are highly sensitive parameters, specific to each dataset, that require intensive tuning. By fixing the step size, the burden is essentially moved from runtime evaluation to offline tuning.

\begin{algorithm}[htbp]
\caption{Batch Gradient Descent (BGD)}
\label{alg:batch-gd}
\algsetup{linenodelimiter=.}

\textbf{Input:} $\{(\vec{x}_{j}, y_{j})\}_{1\leq j\leq N}$, $f$, $\nabla\Lambda$, $\vec{w}^{(0)}$, $\alpha^{(0)}$\\
\textbf{Output:} $\vec{w}^{(k-1)}$

\begin{algorithmic}[1]

\STATE Initialize $\vec{w}^{(0)}$ and $\alpha^{(0)}$ with random values if not provided
\STATE Let $k = 1$

\WHILE {(\textbf{true})}

\STATE \textbf{if} model\_convergence($\{f(\vec{w}^{(l)})\}_{0\leq l < k}$) \textbf{then} \textbf{break}

\STATE Compute gradient: $\nabla \Lambda(\vec{w}^{(k-1)})\{(\vec{x}_{j}, y_{j})\}_{1\leq j\leq N}$
\STATE Update model: $\vec{w}^{(k)} = \vec{w}^{(k-1)} - \alpha^{(k-1)} \nabla \Lambda(\vec{w}^{(k-1)})$
\STATE Update step size $\alpha^{(k)}$

\STATE Let $k = k+1$

\ENDWHILE

\RETURN {$\vec{w}^{(k-1)}$}

\end{algorithmic}

\end{algorithm}

\paragraph*{SQL representation.}
We formulate the batch gradient solution to the objective function in Eq.~(\ref{eq:optim-form}) as SQL aggregate queries. This allows us to identify database-specific optimizations. There are two dataset-wide operations in Algorithm~\ref{alg:batch-gd}---loss evaluation and gradient computation. Both of them can be expressed as SQL queries over a relation $T\left(\vec{x}, y\right)$ in which each tuple represents a training example. The loss at a given point $\vec{w}$ is evaluated by the query:
\begin{sql}
SELECT SUM($f\left(\vec{w}, \vec{x}; y\right)$) FROM T
\end{sql}
in which the vector operations involving the example feature vector $\vec{x}$ and the multi-dimensional point $\vec{w}$ can either be expressed as array functions or be mapped explicitly into arithmetic operators. Gradient computation at a point $\vec{w}$ requires one aggregate for every dimension, as shown in the following SQL query:
\begin{sql}
SELECT SUM($\frac{\partial{f}}{\partial{w_{1}}}\left(\vec{w}, \vec{x}; y\right)$), $\dots$, SUM($\frac{\partial{f}}{\partial{w_{d}}}\left(\vec{w}, \vec{x}; y\right)$)
FROM T
\end{sql}
It is important to emphasize that both the point $\vec{w}$ and the function $f$ and its gradient $\nabla{f}$ are constants at a given iteration in the two queries given above. To make this point clear, we show the actual queries corresponding to SVM classification with -1/+1 labels:
\begin{sql}
SELECT SUM($\sum_{i=1}^{d}{1-yx_{i}w_{i}}$) FROM T
WHERE $\left(\sum_{i=1}^{d}{1-yx_{i}w_{i}}\right)$ > 0
SELECT SUM($-yx_{1}$), $\dots$, SUM($-yx_{d}$) FROM T
WHERE $\left(\sum_{i=1}^{d}{1-yx_{i}w_{i}}\right)$ > 0
\end{sql}
For other model types, only the formula of the loss function and its gradient change. The rest stays the same. Most importantly, the query shape is unchanged.

Based on these examples, we formalize the BGD computations with the following abstract SQL query:
\begin{equation}\label{eq:abstract-query}
\texttt{SELECT SUM(f}_{\texttt{1}}\texttt{(t)), \dots, SUM(f}_{\texttt{p}}\texttt{(t)) FROM T}
\end{equation}
in which $p$ different aggregates are computed over the training tuples in relation $T$. At least two instances of this query have to be executed per iteration---one for the gradient and one for the loss. The exact number is typically much larger and is determined by the number of times the loss is computed in line search.

\eat{
In the parallel solution we propose in Section~\ref{ssec:par-grad-descent:batch}, only one query instance is executed per iteration, albeit with a larger number of aggregates, i.e., $s \cdot (d+1)$ instead of $(d+1)$. Moreover, in order to speed-up the execution of query~(\ref{eq:abstract-query}) further, relation $T$ is partitioned into $M$ blocks that are processed concurrently. The partial sums corresponding to each block are then added together to produce the final result. This solution works because addition is commutative and associative, i.e., it is trivially parallelizable. A factor of $M$ speedup is thus obtained.
}

\subsection{Incremental Gradient Descent}\label{ssec:grad-descent:stochastic}

The problem with BGD is that one pass -- or many more if line search is executed to find the optimal step size -- over the entire data is required in order to move a single step. This is a big issue in our target scenario -- massive number of examples $N$ -- since the time per iteration is linear in $N$ and many iterations are typically required in order to achieve convergence. Incremental or stochastic gradient descent (IGD)~\cite{bertsekas:igd} addresses this issue by taking $N$ steps per iteration. The direction of each step is given by the gradient corresponding to a single data example $\vec{x}_{i}$. Essentially, the entire gradient is approximated with a single term in the summation, i.e., $\nabla \Lambda(\vec{w}) \approx \nabla f(\vec{w}, \vec{x_{i}}; y_{i})$. If we ignore the regularizer $R$ the step recurrence becomes:
\begin{equation}\label{eq:stoch-grad-step}
\vec{w}^{(k+1)} = \vec{w}^{(k)} - \alpha^{(k)} \nabla f \left( \vec{w}^{(k)}, \vec{x}_{\eta^{(k)}}; y_{\eta^{(k)}} \right)
\end{equation}
where $\eta^{(k)}$ returns the $\text{k}^{\text{th}}$ example in a random permutation of the data. The permutation is necessary to guarantee that progress towards convergence is made inside an iteration. Moreover, a different permutation should be used at each iteration to avoid stalling at a non-minimum point and increase the convergence rate. Given the large number of steps taken inside a single iteration, the step size $\alpha^{(k)}$ has to be carefully tuned to minimize extreme oscillations. Executing line search after every step is out of the question. Algorithm~\ref{alg:sgd} summarizes the IGD differences compared to BGD. Lines 5 and 6 in Algorithm~\ref{alg:batch-gd} are replaced with the \texttt{for} loop given in Algorithm~\ref{alg:sgd}, where $\vec{w}^{(k)}_{(0)} = \vec{w}^{(k-1)}$ and $\vec{w}^{(k)} = \vec{w}^{(k)}_{(N)}$, respectively.

\begin{algorithm}[htbp]
\caption{Incremental Gradient Descent (IGD)}
\label{alg:sgd}
\algsetup{linenodelimiter=.}

\begin{algorithmic}[1]
\FOR {$i=1$ \textbf{to} $N$}
\STATE Approximate gradient: $\nabla f \left( \vec{w}^{(k)}_{(i-1)}, \vec{x}_{\eta^{(i)}}; y_{\eta^{(i)}} \right)$
\STATE $\vec{w}^{(k)}_{(i)} = \vec{w}^{(k)}_{(i-1)} - \alpha^{(k)} \nabla f \left( \vec{w}^{(k)}_{(i-1)}, \vec{x}_{\eta^{(i)}}; y_{\eta^{(i)}} \right)$
\ENDFOR
\end{algorithmic}

\end{algorithm}

\paragraph*{Mini-batch gradient descent.}
BGD takes a single step per iteration. IGD takes one step for every data example. An intermediate solution is to estimate the gradient using more than a single term in the summation Eq.~(\ref{eq:optim-form}) and take one step for each group of terms. If there are $n$, $1<n<N$, such groups then $n$ steps are taken in a single iteration. The step size $\alpha$ has to be configured accordingly. This alternative is known as mini-batch gradient descent. While it has been shown that there are situations when the mini-batch solution outperforms both the batch and the incremental methods, the addition of another parameter -- the size of a batch or the number of batches -- only increases the complexity of tuning. This cannot be ignored when the number of data examples $N$ is large.

\paragraph*{SQL representation.}
The main difference between BGD and IGD with respect to SQL representation is that $\vec{w}$ is not constant for an entire iteration. In pure IGD, $\vec{w}$ is updated for every example in the input, while in mini-batch gradient descent, $\vec{w}$ is updated after a fixed number of examples. This can be expressed in SQL as the following \texttt{UPDATE} statement:
\begin{sql}
UPDATE W
SET $w_{1}=w_{1}-\alpha^{(k)} \cdot \frac{\partial{f}}{\partial{w_{1}}}\left(\vec{w}, \vec{x}; y\right)$, $\dots$,
\hspace*{0.75cm}$w_{d}=w_{d}-\alpha^{(k)} \cdot \frac{\partial{f}}{\partial{w_{d}}}\left(\vec{w}, \vec{x}; y\right)$
\end{sql}
where $W(w_{1}, \dots, w_{d})$ is a table with a single tuple containing the model. The pair $\left(\vec{x}; y\right)$ is a tuple from table $T$ corresponding to an input example. The update is executed for all the tuples in $T$, extracted by a cursor in random order. As a concrete example, for SVM classification we have the following SQL statement:
\begin{sql}
UPDATE W
SET $w_{1}=w_{1}-\alpha^{(k)}\cdot yx_{1}$, $\dots$, $w_{d}=w_{d}-\alpha^{(k)}\cdot yx_{d}$
\end{sql}
The loss is computed for the last $\vec{w}$ generated at the end of an iteration, using the same SQL query as in BGD.

\subsection{Beyond First-Order Gradient Descent}\label{ssec:grad-descent:beyond-gd}

Apart from the first-order gradient descent methods presented in this section, coordinate descent~\cite{graphlab} and L-BFGS~\cite{vowpal-wabbit,google-brain} are two alternatives that build upon BGD and IGD.
\textit{Coordinate Descent (CD)} optimizes a multi-dimensional function by minimizing it along one dimension at a time. Instead of moving along the overall gradient direction, CD takes steps along each coordinate direction sequentially. The search along each coordinate is done by line search, which requires a step size.
\textit{L-BFGS} is a quasi-Newton method that searches the model space by approximating the inverse Hessian matrix of the objective function. Different from gradient descent, L-BFGS maintains a history of the last $m$ updates to model $w$ and gradient $\nabla\Lambda$. The direction $\vec{d}$ at iteration $k$ is given by $\vec{d}^{(k)}=-H^{(k)} \nabla\Lambda^{(k)}$, where $H^{(k)}$ is the approximation to the Hessian. Without going into details on how $H^{(k)}$ is computed from historical models and gradients, we point out that, after computing the search direction $\vec{d}^{(k)}$, a line search is performed.

The techniques we propose in this paper can be applied to any of the gradient-based optimization methods -- first- and second-order -- used for large scale model calibration. We emphasize that the number of tunable hyper-parameters, i.e., parameters of the optimization method, is small in all these methods. In the case of coordinate descent, only the step size has to be tuned. Or, perhaps, the number of coordinates to update at the same time, if parallel coordinate descent is conducted. In L-BFGS, only the size of the history and the step size used in line search have to be calibrated. The relatively small space of tunable hyper-parameters allows for more opportunities to apply the techniques proposed in this paper.

\subsection{Discussion}\label{ssec:grad-descent:discuss}

Considering the alternative gradient descent methods discussed in this section, a comparison from the perspective of our specific problem, i.e., Eq.~(\ref{eq:optim-form}) with the assumption that $N$ is extremely large, is required to clarify the merits of each approach. According to the thorough survey by Bertsekas~\cite{bertsekas:igd}, two cases have to be considered. Far from the minimum, IGD can have a convergence rate as much as $N$ times faster for identical loss functions $f$ and large $N$. Close to the minimum, IGD requires diminishing step sizes in order to converge. This results in sub-linear convergence rate---slower than the linear convergence of the batch method. Based on these observations, a hybrid approach~\cite{vowpal-wabbit}, that first executes incremental gradient followed by batch gradient when no progress is made anymore, is likely to be the optimal solution in practice. It is also important to remark that IGD requires data randomization at each iteration in order to achieve the $N$ factor in convergence speedup. With the same cyclical order used across all iterations, IGD does not provide any theoretical convergence improvement beyond BGD~\cite{bertsekas:igd}.

\section{Distributed Gradient Descent}\label{sec:par-grad-descent}

\eat{
In this section, we present existing algorithms to implement gradient descent optimization in parallel and propose novel pipelining strategies to further increase the degree of parallelism. As pointed out by DeWitt and Gray in~\cite{dewitt-paralleldb}, there are two strategies for parallel data processing---data partitioning parallelism and pipelining parallelism. In data partitioning parallelism, data are split into multiple blocks that are processed concurrently either on the same machine or across different machines. In pipelining parallelism, a complex job is first split into a series of independent tasks. The execution of the tasks is then overlapped across different data partitions. It is important to remark that pipelining parallelism requires data partitioning as a prerequisite. While data partitioning is heavily used in existing parallel gradient descent algorithms~\cite{MR-ML-multicore,parallel-igd-multicore,parallel-bgd,parallel-igd,bismarck,vowpal-wabbit}, pipelining is hardly considered. Our contribution is to \textit{investigate if and how pipelining can be used in gradient descent}.
}

In this section, we present parallel algorithms for distributed gradient descent optimization. With almost no exception, all the existing algorithms~\cite{MR-ML-multicore,parallel-igd-multicore,parallel-bgd,parallel-igd,bismarck,vowpal-wabbit} make exclusive use of data partitioning parallelism, i.e., the $M$ subsets are processed concurrently across the execution nodes and the partial results are merged together at the end of an iteration. To the best of our knowledge, other types of parallelism~\cite{dewitt-paralleldb}, e.g., pipelining, super-scalar execution, and vectorization, are not considered in the literature.

\eat{
\textbf{Setup.}
Our goal remains minimizing the objective function in Eq.~(\ref{eq:optim-form}) using gradient descent methods. The training dataset consisting of $N$ (vector, label) pairs $\{(\vec{x}_{1}, y_{1}), \dots, (\vec{x}_{N}, y_{N})\}$ is partitioned into $M$ subsets based on a partitioning function. Each subset is assigned to a different processing node for execution. Following the standard database approach, we assume that the subset corresponding to each node is larger than the available memory, thus a further partitioning into chunks that fit in memory is required. While this data partitioning scheme allows for concurrent data processing across nodes and corresponding speedup, it also impacts the execution and convergence of gradient descent. In addition to processing concurrently the $M$ subsets and multiple chunks on a node, we also consider how to parallelize the intra-iteration execution of the gradient descent methods introduced in Section~\ref{sec:grad-descent}.
}

\subsection{Distributed BGD}\label{ssec:par-grad-descent:batch}

The straightforward strategy to parallelize BGD is to overlap gradient computation (line 5 in Algorithm~\ref{alg:batch-gd}) across the processing nodes storing the $M$ data subsets~\cite{MR-ML-multicore}. The partial gradients are subsequently aggregated at a coordinator node holding the current model $\vec{w}^{(k)}$, where a single global update step is performed in order to generate the new model $\vec{w}^{(k+1)}$ based on Eq.~(\ref{eq:grad-step}). The update step requires completion of partial gradient computation across all the nodes, i.e., update model is a synchronization barrier. Once the new model $\vec{w}^{(k+1)}$ is computed -- using line search or a fixed step size -- it is disseminated to the processing nodes for a new iteration over the data. Parallel BGD works because the objective function $\Lambda$ is linearly separable, i.e., the gradient of a sum is the sum of the gradient applied to each term: $\nabla \left[\sum_{i=1}^{N} f\left(\vec{w}, \vec{x_{i}}; y_{i}\right)\right] = \sum_{i=1}^{N} \nabla f\left(\vec{w}, \vec{x_{i}}; y_{i}\right)$. It provides linear processing speedup and logarithmic communication in the number of nodes since aggregation can be executed on a tree structure. In terms of convergence though, there is no improvement with respect to the sequential algorithm---only faster iterations for gradient computation.

An alternative parallel algorithm is given in~\cite{parallel-bgd}. Each node solves to convergence an independent BGD on its corresponding data subset. The joint model is produced by averaging the partial models computed at each node. As a result, communication between nodes is reduced to minimum. From a convergence standpoint, this ensemble solution provides only variance reduction, but does not eliminate the bias when compared to the single node sub-sample solution~\cite{parallel-igd}. The execution time is dominated by the convergence time of the slowest node.

\eat{
\textbf{How to mitigate the slow node problem?}
Since model update (line 6 in Algorithm~\ref{alg:batch-gd}) is a synchronize-all barrier, it can be executed only after partial gradient computation finishes across all the nodes and the complete gradient is aggregated. Or gradient computation converges for each sub-sample. This represents a problem when there is variation in processing speed across nodes since a slow node can delay the entire computation. The solution proposed in~\cite{VW:arxiv} to mitigate the slow node problem is speculative execution. The same partial gradient is computed in parallel on multiple nodes and the first one to finish is used in aggregation. The others are discarded. It is important to notice that speculative execution requires slow node runtime detection and data movement or data replication. A different approach is to time out the computation after a fixed time interval. Incomplete results -- gradients or models -- from slow nodes can be either included in the final result computation or discarded altogether. It is not clear which alternative is optimal in all the situations. At a high-level, this approach is similar to the mini-batch solution since the gradient is approximated from a subset of the terms. The main difference is that a single step is still taken for every iteration over the data.
}

\subsection{Distributed IGD}\label{ssec:par-grad-descent:stochastic}

Intuitively, parallelizing IGD cannot be any different from the batch version since the difference between the two algorithms is minimal. It should be even simpler since no global gradient computation is required. Essentially, a model update is executed for every data example based solely on its value. At a closer look though, we observe that the updates in Algorithm~\ref{alg:sgd} are strictly sequential, i.e., the input of the current update is the output of the previous one. Moreover, the order in which updates are executed affects the final result, i.e., model update is not commutative. These reasons seem to preclude any type of parallelism altogether. Unless we drop the requirement that all the data examples have to be used to update the model in one iteration. Similar to the approach in~\cite{parallel-bgd}, the standard method to parallelize IGD~\cite{parallel-igd} is to solve a separate model for every data partition and then to average the $M$ models in order to generate the final model at each iteration. The resulting model is then passed as a starting point to the subsequent iteration and the process is repeated until convergence, which is theoretically achieved in a logarithmic number of iterations~\cite{parallel-igd}. An alternative that minimizes communication is to execute IGD until convergence on each partition and average only the final models.

\paragraph*{Model averaging.}
The following tradeoff has to be considered in the case of parallel IGD. When the number of data partitions $M$ is large, so is the degree of parallelism. This results in higher speedup. It also results in more models computed over fewer examples to be averaged. This is a potential problem since there is a lower bound on the number of examples required for a given model to achieve the asymptotic regime~\cite{parallel-igd}. Moreover, the examples seen by every model have to represent a random sample from the entire population. The larger the number of partitions $M$, the higher the communication required to generate random samples. The combined effect of non-asymptotic behavior and non-random examples is higher variance across the partial models, which finally results in slower convergence~\cite{bismarck,igd-glade}. Thus, choosing the optimal $M$ value has deeper implications in the case of parallel IGD.

\paragraph*{Shared-memory parallelization.}
Exactly because of this reason, the preferred solution to parallelize IGD in a shared-memory setting is to discard model merging completely. A single model is shared across all the threads, while concurrent updates to the model are serialized through locking. Unfortunately, this solution reduces speedup dramatically even when light locking mechanisms, e.g., \texttt{compare\&swap}, are used. The solution proposed in~\cite{nolock-igd} to solve this problem is to discard locking altogether and allow for contentious updates to the shared model. As expected, this results in linear speedup. What about convergence? As long as the model is highly-dimensional and the gradients used in updates are sparse, i.e., only a small number of model dimensions are updated, the difference in convergence compared to the serial solution is minimal.

\paragraph*{Hybrid solution.}
In a distributed setting with multi-core nodes, a hybrid approach is regarded as the optimal solution~\cite{lmf-hybrid-glade}. A single model is created for each node. It is shared across all the local threads or processes and updated using one of the shared-memory strategies, chosen based on the actual model properties. Model merging across nodes is executed using the standard averaging technique along an aggregation tree. The hybrid solution aims to maximize the degree of parallelism available in the system without dramatically impacting the convergence.

\subsection{Summary}\label{ssec:par-grad-descent:discussion}

\eat{
Parallelizing gradient descent methods in a distributed environment poses different challenges. For the batch method, gradient computation is trivially parallelizable due to the objective function being linearly separable. The main difficulty is posed by model update which is a strictly sequential operation. We propose a SIMD approach in which multiple updates are computed concurrently at virtually no cost. Only the optimal model is kept for further processing. Moreover, we overlap loss and gradient computation, thus cutting the number of passes over the data in half. The rationale behind this approach is that data access is the bottleneck in the out-of-memory target scenario we consider and plenty of parallelism is available in the current multi-core CPUs---not to mention GPUs. The main issue when parallelizing incremental gradient descent is the tradeoff between the degree of parallelism and the convergence rate. More parallelism does not automatically result in better convergence. We argue for a hybrid approach in which a shared model is created for each individual node while averaging is employed to merge models across nodes.
}

Parallelizing gradient descent methods in a distributed environment poses different challenges. For the batch method, gradient computation is trivially parallelizable due to the objective function being linearly separable. The main difficulty is posed by model update which is a strictly sequential operation. The main issue when parallelizing IGD is the tradeoff between the degree of parallelism and the convergence rate. More parallelism does not automatically result in better convergence. A hybrid approach in which a shared model is created for each node, while averaging is used to merge models across nodes, is likely to be the optimal solution in general.

\section{Speculative Iterations}\label{sec:multiple-steps}

In this section, we address two fundamental problems specific to model calibration---convergence detection and parameter tuning. As with any iterative method, gradient descent convergence is achieved when there is no more decrease in the objective function, i.e., the loss, across consecutive iterations. While it is obvious that convergence detection requires loss evaluation at every iteration, the standard practice, e.g., Vowpal Wabbit~\cite{vowpal-wabbit}, MLLib~\cite{mllib}, is to discard detection altogether and execute the algorithm for a fixed number of iterations. The reason is simple: loss computation requires a complete pass over the data, which doubles the execution time. This approach suffers from at least two problems. First, it is impossible to detect convergence before the specified number of iterations finishes. And second, it is impossible to identify bad parameter configurations, i.e., configurations that do not lead to model convergence. Recall that both BGD and IGD depend on a series of parameters, the most important of which is the step size. Finding the optimal step size typically requires many trials. Discarding loss computation increases both the number of trials as well as the duration of each trial.

\paragraph*{High-level approach.}
We propose a unified solution for convergence detection and parameter tuning based on speculative processing. The main idea is to \textit{overlap gradient and loss computation for multiple parameter configurations across every data traversal}. This allows for timely convergence detection and early bad configuration identification since many trials are executed simultaneously. Overall, faster model training. The intuition behind our approach is that modern CPU architectures provide extensive parallelization opportunities, e.g., multi-core, hyper-threading, vectorization, that are difficult to use at full potential in disk-based workloads specific to Big Data analytics over massive datasets. With the right mix of tasks and judicious scheduling, the degree of parallelism supported in hardware can be fully utilized by executing multiple tasks concurrently. Moreover, the overall execution time is similar to the time it takes to execute each task separately.

\begin{figure*}[t]
\begin{center}
\includegraphics[width=0.9\textwidth]{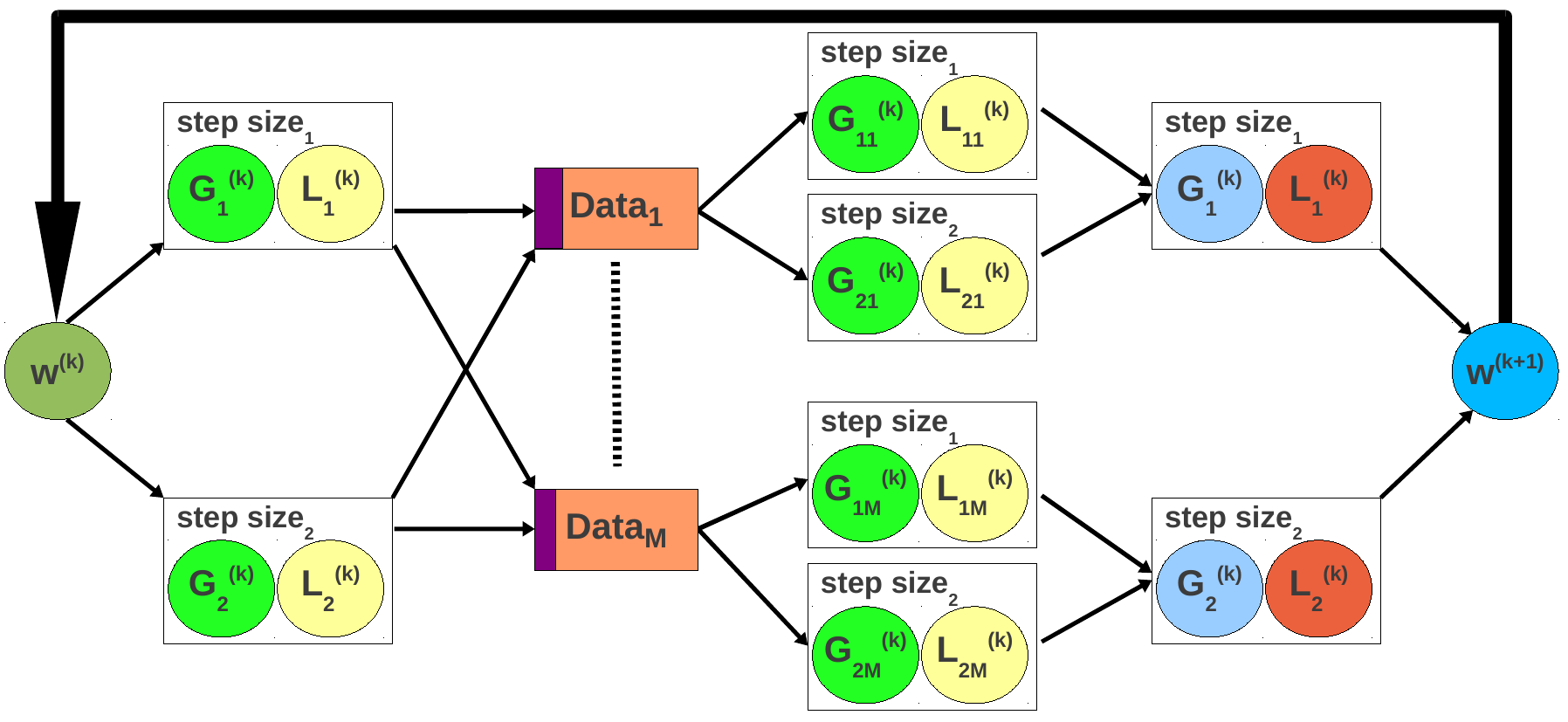}
\caption{Speculative parameter testing with online aggregation.}
\label{fig:architecture}
\end{center}
\end{figure*}

Our contribution is to \textit{design speculative gradient descent algorithms that test multiple step sizes simultaneously and overlap gradient and loss computation}. The number of step sizes used at each iteration is determined adaptively and dynamically at runtime. The step sizes are drawn from a parametric distribution that is continuously updated using a Bayesian statistics~\cite{bayes-stat} approach. Only the model with the minimum loss survives each iteration, while the others are discarded (Figure~\ref{fig:architecture}). As long as the execution time does not dramatically increase when multiple step sizes are tested, speculative execution leads to faster model training. To the best of our knowledge, this is the first solution that uses speculative execution for gradient descent parameter tuning and loss computation. Speculative execution is applied in Vowpal Wabbit~\cite{vowpal-wabbit} and MLlib~\cite{mllib}, but only to deal with the problem of slow nodes. Not to test multiple step sizes simultaneously. Vowpal Wabbit also supports progressive loss estimation which overlaps gradient and loss computation. Progressive loss provides only local estimates. It never computes the exact loss at the end of an iteration. This can be done only with an additional pass over the data. Our solution is exact -- not approximate -- and works in a distributed setting.

\subsection{Speculative BGD}\label{ssec:spec-grad-descent:batch}

\begin{algorithm}[htbp]
\caption{Speculative BGD}
\label{alg:speculative-bgd}
\algsetup{linenodelimiter=.}

\textbf{Input:} $\{(\vec{x}_{j}, y_{j})\}_{1\leq j\leq N}$, $f$, $\nabla\Lambda$, $\vec{w}^{(0)}$, $\nabla\Lambda(\vec{w}^{(0)})$, $s$, $\Phi$\\
\textbf{Output:} $\vec{w}^{(k)}$

\begin{algorithmic}[1]
\STATE Initialize $\vec{w}^{(0)}$ and $\nabla\Lambda(\vec{w}^{(0)})$ randomly if not provided
\STATE Let $k=1$

\WHILE {(\textbf{true})}

\STATE Draw $s$ step sizes $\{\alpha_{1}, \dots, \alpha_{s}\}$ from distribution $\Phi$
\STATE Let $\vec{w}_{i} = \vec{w}^{(k-1)} - \alpha_{i} \nabla \Lambda(\vec{w}^{(k-1)})$, $1 \leq i \leq s$

\FOR {\textbf{each example} $(\vec{x}_{j}, y_{j})$}
\STATE \textbf{for each model} $\vec{w}_{i}$ \textbf{do \textit{in parallel}}
\STATE \hspace*{0.25cm} Compute gradient: $\nabla \Lambda(\vec{w}_{i}) \{(\vec{x}_{j}, y_{j})\}$
\STATE \hspace*{0.25cm} Compute loss: $f(\vec{w}_{i}, \vec{x}_{j}; y_{j})$
\STATE \textbf{end for}
\ENDFOR

\STATE Let $\vec{w}^{(k)}=\textit{min}_{f(\vec{w}_{i})} \{\vec{w}_{i}\}$, $1 \leq i \leq s$
\STATE \textbf{if} model\_convergence($\{f(\vec{w}^{(l)})\}_{0\leq l < k}$) \textbf{then} \textbf{break}

\STATE Update step distrib: $\Phi=\textit{Bayes}(\Phi, \{\alpha_{i}\}, \{f(\vec{w}_{i})\})_{i \leq s}$
\STATE Update number of steps $s$ based on nested loops (lines 6-11) execution time

\STATE Let $k = k+1$

\ENDWHILE

\RETURN {$\vec{w}^{(k)}$}

\end{algorithmic}

\end{algorithm}

Speculative BGD works as follows. Define a set of possible step sizes $\{\alpha_{1}, \dots, \alpha_{s}\}$, $s \geq 1$. Generate an updated model $\vec{w}_{i}^{(k+1)}$, $1 \leq i \leq s$, for each of these step sizes and compute the corresponding objective function value concurrently. Choose the model $\vec{w}_{i}^{(k+1)}$ with the minimum loss as the new model. Repeat the procedure for every iteration until convergence. The pseudo-code for the speculative BGD algorithm is given in Algorithm~\ref{alg:speculative-bgd}. The statements contained inside an \textit{in parallel} loop are executed concurrently across iterations. Moreover, in the case of \textit{Speculative BGD}, the two statements inside the loop (lines 8 and 9) are executed in parallel even for the same iteration. In the following, we discuss the main components of the algorithm.

\eat{
Recall that in our target scenario disk access is the bottleneck since the number of examples $N$, respectively $M$, is considerably larger than the available memory. Thus, as long as the increased computation volume does not offset the time taken by multiple synchronized iterations over the data, the proposed strategy has the chance to outperform line search methods in running time.
}

\paragraph*{Concurrent step size evaluation.}
Intuitively, speculative processing replaces multiple passes over the data with a single pass that requires $\mathcal{O}(s)$ times more computation. In SQL terms, this corresponds to merging $s$ independent aggregate queries into a single query with $\mathcal{O}(s)$ aggregates. This provides a significant reduction in execution time even for traditional I/O-bound databases. Modern databases that take advantage of state-of-the-art parallel hardware architectures can do much better and achieve similar execution times as for a single aggregate query. There are two levels of parallelism that can be exploited---multi-threaded execution on multi-core CPUs and vectorized instructions. Instead of updating all the $s$ models in a single thread, each model is updated in a separate thread, initialized with a different model. All the threads perform exactly the same processing over the same example data. Only their initialization models are different. This is a standard instance of SIMD parallelism that is directly supported by multi-core CPUs. Given that every example has to be processed by all the threads, scheduling example access requires discussion. One approach is to make every example available to all the threads. While the threads can process the example concurrently, since read-only access is sufficient, synchronization is still required in order to determine when all the threads finish processing. Another approach is to organize the threads into a pipeline. This allows for overlapped execution across examples, i.e., each thread processes a different example, and eliminates barrier synchronization. Vectorized processing, e.g., \texttt{SSE} and \texttt{AVX} instructions, can be applied in two places. A set of examples can be grouped together into a chunk and processed as a unit inside a thread. As a side effect, chunking also reduces the overhead of moving examples through the pipeline. The second instance for vectorization is dot-product computation, which represents the primitive operation in gradient and loss evaluation.

\paragraph*{Gradient and loss computation overlapping.}
The model that minimizes the loss across the $s$ possible step sizes is taken as input by the subsequent iteration. The gradient has to be computed for this model. Remember though that we have already used this model to compute the loss in the previous iteration. Unfortunately, we did not know if this model is selected or not. Under the assumption that data access dominates the execution time, we argue for bundling loss and gradient computation together in order to save one pass over the data. Overall, the number of data traversals is determined entirely by the number of gradient computations while loss evaluation piggybacks on the data access. We have to apply the overlapping strategy for all the $s$ models though, since we do not know the one minimizing the loss. Although we might expect this to increase the execution time of an iteration, the aggressive use of SIMD pipeline parallelism across the $s$ step sizes minimizes the impact of additional computation and higher memory usage, as the experimental results in Section~\ref{sec:experiments} show.

\paragraph*{How many step sizes $s$?}
The larger the number of step sizes, the higher the probability to find a model with smaller loss. However, a too large degree of speculation might result in an unacceptable increase in iteration time. Thus, choosing the right number of step sizes is important for optimal speculative processing. What makes the problem even more difficult is the high variance in processing time as a function of model type and size. We propose an adaptive solution that selects the number of step sizes dynamically at runtime based on available resources. Moreover, the number of step sizes varies from one iteration to another. We start with a single step size in the first iteration and measure the execution time. For each subsequent iteration, we increase the number of step sizes, e.g., exponentially, as long as the increase in execution time remains below a specified threshold. We use this number of steps for all the subsequent iterations. In order to handle resource fluctuations in the system, we continuously monitor the execution time per iteration. In the case of a significant increase in processing time, we reduce the number of steps accordingly.

\paragraph*{How to choose the step sizes?}
The simple solution to choosing the step sizes at a given iteration is to have $s$ constants that cover a large range of values, some of which are very small. The constants can be kept the same across iterations or they can be decreased at some rate, i.e., the decay in IGD. The problem with this approach is that we discard the knowledge we gain from previous iterations. Ideally, we want to set the current $s$ values based on the previous best-performing step sizes. Bayesian statistics~\cite{bayes-stat} provides a principled framework to accomplish this. We start with a prior parametric distribution for the step size. This can be a single distribution or a mixture. The prior can be learned from the historical workload or it can be set to some default distribution, e.g., normal. The $s$ step sizes used at each iteration are sampled randomly from the current distribution. The resulting losses are normalized and converted to probabilities. The $s$ pairs (step size, loss) are combined together with the prior in order to compute the updated step size distribution---known as the posterior. This consists in determining the posterior distribution parameters and it can be done with a straightforward Expectation-Maximization (EM) algorithm that identifies the Maximum Likelihood Estimator (MLE) based on the available samples. We do not provide the details of the EM algorithm since it is dependent on the shape of the prior. The posterior distribution becomes the prior in the subsequent iteration and the entire procedure is repeated.

\eat{
Ideally, we want to find that point along the opposite gradient direction where the objective function value is minimized. By limiting the number of step sizes we consider, the likelihood of finding this point is very low even when the number of sizes $s$ is large. The probability of finding a point with a lower objective value increases though linearly with the number of sizes---especially if these cover a large range of values, some of which are very small. Based on this argument, we are almost guaranteed that progress towards the minimum is made at each iteration. A more elaborated approach is to define a probabilistic distribution over the step sizes. At each iteration, $s$ sizes are sampled from the distribution and used to update the model. The distribution can be learned from historical data and updated at runtime based on a Bayesian process~\cite{bayes-stat} in order to better reflect the current data.
}

\begin{algorithm}[htbp]
\caption{Speculative IGD}
\label{alg:speculative-igd}
\algsetup{linenodelimiter=.}

\textbf{Input:} $\{(\vec{x}_{j}, y_{j})\}_{1\leq j\leq N}$, $f$, $\nabla\Lambda$, $\{\vec{w}^{(0)}_{i}\}_{1\leq i\leq s}$, $\Phi$\\
\textbf{Output:} $\vec{w}^{(k)}$

\begin{algorithmic}[1]
\STATE Initialize $\{\vec{w}^{(0)}_{i}\}_{1\leq i\leq s}$ randomly if not provided
\STATE Let $k=1$

\WHILE {(\textbf{true})}

\STATE Draw $s$ step sizes $\{\alpha_{1}, \dots, \alpha_{s}\}$ from distribution $\Phi$
\STATE Let $\vec{w}_{il}=\vec{w}^{(k-1)}_{i}$, $1\leq i,l\leq s$

\FOR {\textbf{each example} $(\vec{x}_{\eta^{(j)}}, y_{\eta^{(j)}})$}

\STATE \textbf{for each model} $\vec{w}_{il}$ \textbf{do \textit{in parallel}}

\STATE \hspace*{0.25cm} Approximate gradient: $\nabla f \left( \vec{w}_{il}, \vec{x}_{\eta^{(j)}}; y_{\eta^{(j)}} \right)$
\STATE \hspace*{0.25cm} Update: $\vec{w}_{il} = \vec{w}_{il} - \alpha_{l} \nabla f \left( \vec{w}_{il}, \vec{x}_{\eta^{(j)}}; y_{\eta^{(j)}} \right)$

\STATE \textbf{end for}

\STATE \textbf{for each original model} $\vec{w}^{(k-1)}_{i}$ \textbf{do \textit{in parallel}}

\STATE \hspace*{0.25cm} Compute loss: $f \left( \vec{w}^{(k-1)}_{i}, \vec{x}_{\eta^{(j)}}; y_{\eta^{(j)}} \right)$

\ENDFOR

\STATE Let $\vec{w}^{(k)}=\textit{min}_{f(\vec{w}^{(k-1)}_{i})} \{\vec{w}^{(k-1)}_{i}\}$, $1 \leq i \leq s$
\STATE Let $m$ be the index of the minimum loss $f(\vec{w}^{(k-1)}_{i})$

\STATE \textbf{if} model\_convergence($\{f(\vec{w}^{(l)})\}_{0\leq l < k}$) \textbf{then} \textbf{break}

\STATE Update step distrib: $\Phi=\textit{Bayes}(\Phi, \{\alpha_{i}\}, \{f(\vec{w}_{mi})\})_{i \leq s}$
\STATE Update number of steps $s$ based on nested loops (lines 6-13) execution time

\STATE Let $\vec{w}^{(k)}_{i}=\vec{w}_{mi}$, $1 \leq i \leq s$
\STATE Let $k=k+1$

\ENDWHILE
\RETURN {$\vec{w}^{(k)}$}
\end{algorithmic}

\end{algorithm}

\subsection{Speculative IGD}\label{ssec:spec-grad-descent:incremental}

At high level, the speculative techniques proposed for BGD are directly applicable to IGD: compute $s$ models instead of one and overlap model update with loss evaluation. At closer look though, there is a significant difference. While computing the loss for a given model, the model changes continuously since multiple steps are taken during the update phase. As a result, the loss obtained at the end of an iteration corresponds to the starting model, while the final model is the updated model generated from the same starting model. Notice that they are different models though. This creates problems for at least two reasons. First, it is not clear that the resulting model corresponding to the original model having the minimum loss is the optimal model to select. And second, $s$ step sizes for $s$ initial models generate $s^{2}$ resulting models after one iteration. Since the number grows exponentially with the number of iterations, a pruning mechanism that selects only $s$ models at every iteration is required. We address both these issues with the following strategy (Algorithm~\ref{alg:speculative-igd}): select the $s$ resulting models corresponding to the initial model having minimum loss. This guarantees that the starting model is optimally chosen. Since we do not know which of the $s$ step sizes is optimal, we keep the models corresponding to all of them. The models generated for the sub-optimal $(s-1)$ initial models are all discarded. This provides the necessary pruning mechanism to support efficient processing. There are two \textit{in parallel} loops in \textit{Speculative IGD}---one for continuously updating models and one for the original models. While these two loops can be processed concurrently, the loop for continuous updating (lines 8 and 9) is sequential inside the same iteration.

\subsection{Summary}\label{ssec:vector-grad-descent:discussion}

Speculative processing allows for multiple models to be evaluated concurrently by taking advantage of the shared access to the same example data. When this is overlapped with loss computation, the parameter search space can be explored more effectively. We discuss the mechanisms required to support efficient speculative execution for gradient descent optimization parameter tuning. Specifically, we propose adaptive algorithms for choosing the number of step sizes and their values at every iteration. While these algorithms apply to both BGD and IGD, additional problems are introduced by the divergence between the initial model and the resulting models in the case of IGD. We present an heuristic strategy that selects optimal models and prunes the search space to allow for efficient evaluation.

\section{Intra-Iteration Approximation}\label{sec:ola-grad-descent}

The speculative processing methods proposed in Section~\ref{sec:multiple-steps} allow for a more effective exploration of the parameter space. However, they still require complete passes over the entire data at each iteration in order to detect the sub-optimal parameter configurations. A complete pass is often not required in the case of massive datasets due to redundancy. It is quite likely that a small random sample summarizes the most representative characteristics of the dataset and allows for the identification of the sub-optimal configurations much earlier. This results in tremendous resource savings, more focused exploration, and faster convergence.

\paragraph*{High-level approach.}
We present a \textit{novel solution for using online aggregation sampling in parallel gradient descent optimization to speed-up the execution of a speculative iteration}. We generate samples with progressively larger sizes dynamically at runtime and execute gradient descent optimization incrementally, until the approximation error drops below a user-defined threshold $\epsilon$. Relative to Figure~\ref{fig:architecture}, the entire process is executed multiple times during an iteration, on samples with increasing size. This is completely different from static sub-sampling methods that first extract a fixed-size random sample and then execute gradient descent on the sample, expecting that the optimal solution over the sample is also optimal for the entire dataset. Moreover, online aggregation avoids the expensive re-sampling required in sub-sampling whenever the approximation error level is not satisfied. While parallel online aggregation has been studied before for SQL aggregates~\cite{par-hash-ripple,distributed-ola,MR-ola,hadoop-online,blink,pfola:dapd,agarwal:bootstrap,zeng:bootstrap}, we are the first to consider online aggregation for complex analytics such as gradient descent optimization.

The most important challenge we have to address is how to \textit{design and compute multiple concurrent sampling estimators} corresponding to speculative gradient and loss computation. These estimators arise in the components of multi-dimensional gradients and in the speculative objective function evaluation. Our main contributions can be summarized as follows. We formalize gradient descent optimization as aggregate estimation. We provide efficient parallel solutions for the evaluation of the speculative estimators and of their corresponding confidence bounds that guarantee fast convergence. We design halting mechanisms that allow for the speculative query execution to be stopped as early as the user-defined accuracy threshold $\epsilon$ is achieved.

\subsection{Approximate BGD}\label{ssec:ola-grad-descent:batch}

Exact evaluation of query~(\ref{eq:abstract-query}) can be a lengthy process when the size of data $|T|$ is large or when the computation of a particular aggregate $f_{i}$ is complicated even when parallel solutions are used. The only alternative to speed-up computation further is to resort to approximation. Instead of computing the aggregates exactly, they are only estimated. The estimators have to come with sound accuracy guarantees though in order to guarantee verifiable results. Out of the many approximation techniques proposed in the literature, we argue that sampling is best suited for the aggregate estimation in query~(\ref{eq:abstract-query}). This is because other methods, e.g., sketches~\cite{aqp-book}, have to be constructed separately for every aggregate. This is infeasible in gradient descent optimization since a different set of aggregates have to be computed at each iteration. Only sampling maintains the identity of the data items and supports the computation of any aggregate as it would be computed over the entire data.

\subsubsection{Sampling-Based Estimation}\label{sssec:est:sampling}

Sampling works as follows. A small dataset $T'$ is randomly sampled from $T$.  Query~(\ref{eq:abstract-query}) is executed over the sample $T'$ and the result is scaled-up to compensate for the difference in size between $T$ and $T'$. It is straightforward to show that $\frac{|T|}{|T'|}\cdot{Z_{f_{i}}}$ is an unbiased estimator for the summation corresponding to function $f_{i}$, where $Z_{f_{i}}$ is the result of query~(\ref{eq:abstract-query}) executed over $T'$. Moreover, the accuracy, i.e., confidence bounds, of the estimator can be derived by estimating the variance over the same random sample $T'$. All this requires is the addition of another summation, i.e., $\texttt{SUM(f}^{\texttt{2}}_{\texttt{i}}\texttt{(t))}$, to query~(\ref{eq:abstract-query}) for every aggregate function. While this is standard sampling-based single aggregate estimation treated extensively both in statistics as well as in databases~\cite{aqp-book}, what is specific to our problem is the concurrent computation of multiple aggregates. As far as we know, this is a novel problem that has not been considered extensively in the literature. We discuss aspects specific to the concurrent sampling-based estimation of multiple aggregates in the following.

\paragraph*{How to avoid correlation between estimators?}
The same sample $T'$ is used to estimate all the aggregates in query~(\ref{eq:abstract-query}). In order to avoid correlations between estimators, care has to be taken when extracting the sample. For example, if there is correlation between two dimensions $x_{i}$ and $x_{j}$ in the feature vector, e.g., $x_{i}$ functionally determines $x_{j}$, sampling based on $x_{i}$ determines the values of $x_{j}$, thus it affects the accuracy of the estimators containing $x_{j}$. This type of issues can be avoided if sampling takes into consideration the value of either all the tuple attributes or none of them. For example, a Bernoulli sampling process in which a coin is tossed independently for every tuple does not depend on any of the attribute values. A random hash function that takes as parameter an entire tuple falls in the other category. Both choices are guaranteed to avoid correlation between the aggregate estimators.

\paragraph*{How to choose the optimal sample size $|T'|$?}
To guarantee a given accuracy, the sample size has to be above a certain threshold. Since the threshold is dependent on the actual estimator and its corresponding variance, it cannot be computed directly. It can only be estimated from the historical workload. If the chosen sample size is too small and cannot guarantee the required accuracy, a larger sample has to be extracted from the data. This is a complicated process that takes significant time, e.g., $\mathcal{O}(|T|\log{|T|})$ for sampling without replacement, and requires random disk access. The situation is even more complicated for our particular problem since there are $p$ estimators in query~(\ref{eq:abstract-query}) and the formula of these estimators is different from one iteration to another. Thus, it is likely that a sample taken a priori is not able to guarantee the required accuracy, unless it is considerably large.

\begin{algorithm}[htbp]
\caption{Approximate BGD}
\label{alg:approximate-bgd}
\algsetup{linenodelimiter=.}

\begin{algorithmic}[1]

\FOR {\textbf{each example} $(\vec{x}_{\eta^{(j)}}, y_{\eta^{(j)}})$}
\STATE \textbf{for each \textit{active} model} $\vec{w}_{i}$ \textbf{do \textit{in parallel}}
\STATE \hspace*{0.25cm} Estimate gradient $G_{i}[\textit{est}, \textit{std}]$: $\nabla \Lambda(\vec{w}_{i})\{(\vec{x}_{\eta^{(j)}}, y_{\eta^{(j)}})\}$
\STATE \hspace*{0.25cm} Estimate loss $L_{i}[\textit{est}, \textit{std}]$: $f(\vec{w}_{i}, \vec{x}_{\eta^{(j)}}; y_{\eta^{(j)}})$
\STATE \textbf{end for}

\STATE \textbf{if} test\_est\_convergence($j$) \textbf{then}
\STATE \hspace*{0.25cm} Prune out models: $\textit{Stop Loss} (\{L_{i}[\textit{est}, \textit{std}]\}_{i\leq s}, .05)$
\STATE \hspace*{0.25cm} Let $t$ be the number of remaining models, i.e., \textit{active}

\STATE \hspace*{0.25cm} \textbf{if} ($t = 1$) \textbf{and} $\textit{Stop Gradient} (\{G_{tl}[\textit{est}, \textit{std}]\}_{l\leq d}, .05)$ \textbf{then} \textbf{break}
\STATE \hspace*{0.25cm} Let $s=t$
\STATE \textbf{end if}

\ENDFOR

\end{algorithmic}

\end{algorithm}

\subsubsection{Online Aggregation}\label{sssec:est:ola}

The main idea in online aggregation (OLA)~\cite{ola} is to sample at runtime during normal query processing. Sampling and estimation are essentially overlapped with query execution. As more data are processed towards computing the final aggregates, the size of the sample increases and the accuracy of the estimators -- as reflected by the width of the confidence bounds -- improves accordingly. Whenever the targeted accuracy is achieved, the query, i.e., gradient iteration, can be stopped and a new iteration can be started. In the worst case, the aggregation is executed over the entire data and the exact results are obtained. Algorithm~\ref{alg:approximate-bgd} depicts the pseudo-code for BGD with online aggregation. We present only the nested loops portion of the algorithm. They replace the corresponding nested loops in \textit{Speculative BGD} (lines 6-11). \texttt{test\_est\_convergence} can be triggered either based on the number of examples processed, or externally.

OLA is beneficial for at least two reasons. First, the sample size does not have to be determined before query execution. This eliminates the need for re-sampling altogether. And second, a different sample can be generated at each iteration by simply changing the order in which the examples are processed. This eliminates correlations between estimators across iterations and improves convergence. The main drawback of OLA is that the same query is executed twice---once for computing the correct result and once for computing the estimator. We argue -- and show experimentally in Section~\ref{sec:experiments} -- that this is not a problem since both queries access the same data.

\paragraph*{How to sample efficiently at runtime?}
According to the literature~\cite{olken-phd}, there are two methods to generate samples from a database at runtime. The first method uses an index that provides the random order in which to access the data. This method is highly inefficient due to the large number of random disk accesses. The second method stores data in random order on disk such that a sequential scan returns random samples. Data randomization is executed as a pre-processing step during loading, thus it is a single time cost. In order to generate different samples at each iteration, we choose the starting block in sequential scanning randomly. This guarantees minimal -- if at all -- interference with the actual query processing. Moreover, larger samples can be obtained by running the scan longer. A full scan inspects all data and computes the exact result.

\begin{algorithm}[htbp]
\caption{Stop Gradient}
\label{alg:stop-gradient}
\algsetup{linenodelimiter=.}

\textbf{Input:} $\{[\textit{estimate}_{i}, \textit{std}_{i}]\}_{i\leq d}$, $\epsilon$

\begin{algorithmic}[1]
\STATE Let $\epsilon' = d\cdot \epsilon$
\STATE Let $\textit{error} = \sum_{i=1}^{d}\frac{(\textit{estimate}_{i}+\textit{std}_{i})-(\textit{estimate}_{i}-\textit{std}_{i})}{\textit{estimate}_{i}} = \sum_{i=1}^{d}\frac{2\cdot \textit{std}_{i}}{\textit{estimate}_{i}}$

\STATE \textbf{if} ($\textit{error} \leq \epsilon'$) \textbf{then} \textbf{return} \textbf{true}
\STATE \textbf{else} \textbf{return} \textbf{false}

\end{algorithmic}

\end{algorithm}

\paragraph*{When to stop gradient computation?}
In the case of gradient computation, there are $d$ aggregates in query~(\ref{eq:abstract-query}), one for every dimension in the feature vector. With online aggregation active, there is an estimator and corresponding confidence bounds for every aggregate. Given a desired level of accuracy, we have to determine when to stop the computation and move to a new iteration. We measure the accuracy of an estimator by the \textit{relative error}, defined as the ratio between the confidence bounds width and the estimate, i.e., $\frac{\textit{high}-\textit{low}}{\textit{estimate}}$. For example, if we aim for 95\% accuracy, the relative error has to be below the threshold $\epsilon = 0.05$. What makes our problem difficult is that we have $d$ independent relative errors, one for each estimator. How do we determine when the desired accuracy level is reached in this situation? The simple solution is to wait until all the errors drop below the threshold $\epsilon$. This might require processing the entire dataset even when only a few estimators do not satisfy the accuracy requirement, thus defeating the purpose of online aggregation. An alternative that eliminates this problem is to require that only a percentage of the estimators, e.g., 90\%, achieve the desired accuracy. Another alternative is to define a single convergence threshold across the $d$ estimators. For example, we can define $\epsilon' = d \cdot \epsilon$ and require that the sum of the relative errors across estimators is below $\epsilon'$ (Algorithm~\ref{alg:stop-gradient}). Note that none of these alternatives outperforms the others in all the cases. While we can quantify the effect of each choice on gradient estimation, it is not clear what is the effect on convergence.

\begin{algorithm}[htbp]
\caption{Stop Loss}
\label{alg:stop-loss}
\algsetup{linenodelimiter=.}

\textbf{Input:} $\{[\textit{estimate}_{i}, \textit{std}_{i}]\}_{i\leq s}$, $\epsilon$\\
\textbf{Output:} estimators with considerable overlap, i.e., overlap $\geq \epsilon$

\begin{algorithmic}[1]

\STATE Let $\textit{low}_{i}=\textit{estimate}_{i}-\textit{std}_{i}$; $\textit{high}_{i}=\textit{estimate}_{i}-\textit{std}_{i}$

\STATE Discard all estimators $j$ s.t. $\exists i\leq s$ with $\textit{high}_{i}\leq \textit{low}_{j}$

\STATE Discard all estimators $j$ s.t. $\exists i\leq s$ with $\textit{high}_{i}\leq \textit{low}_{j} + \epsilon$

\RETURN remaining estimators

\end{algorithmic}

\end{algorithm}

\paragraph*{When to stop loss computation?}
While the estimators in gradient computation are independent -- in the sense that there is no interaction between their confidence bounds with respect to the stopping criterion -- the loss estimators corresponding to different step sizes are dependent. Our goal is to choose only the estimator generating the minimum loss. Whenever we determine this estimator with high accuracy, we can stop the execution and start a new iteration. Notice that finding the actual loss -- or an accurate approximation of it -- is not required if gradient descent is executed for a fixed number of iterations.

\begin{figure}[htbp]
\centering
\includegraphics[width=.8\textwidth]{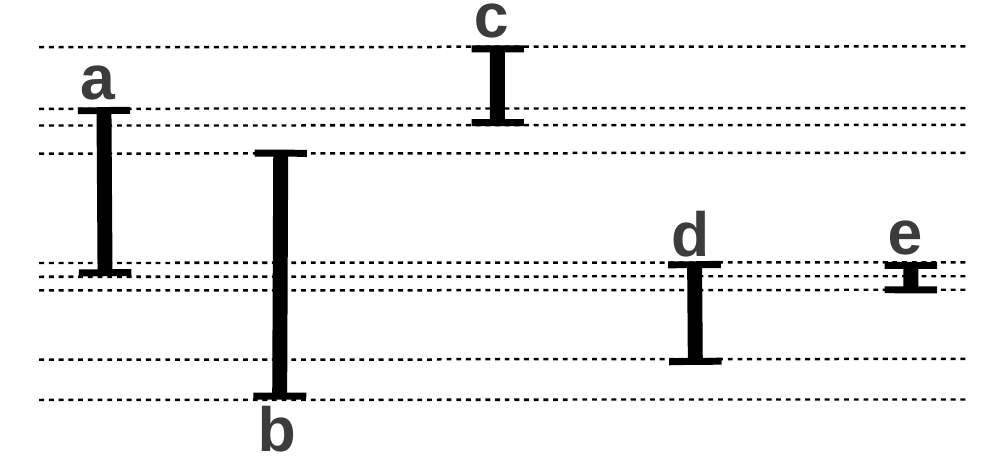}
\caption{Confidence bounds relative relationships.}
\label{fig:ola-loss-stop}
\end{figure}

Figure~\ref{fig:ola-loss-stop} depicts a possible confidence bounds configuration for five loss estimators. Although spread horizontally, this configuration is generated at the same time instant during processing. Lower vertical points correspond to lower loss values. It is clear from the figure that estimator \texttt{c} has no chance to generate the minimum loss, thus it can be discarded. Estimator \texttt{a} overlaps with all of \texttt{b}, \texttt{d}, and \texttt{e}, thus it can still generate the minimum loss. At closer look, we observe that the confidence bounds for \texttt{e} are quite tight and the overlap with \texttt{a} is minimal. Thus, with high probability, \texttt{e} has smaller loss than \texttt{a}, which can be also discarded, without a significant impact on finding the minimum loss. While \texttt{e} is contained entirely inside \texttt{d}, it is at the upper-end of \texttt{d}, i.e., the lower bound of \texttt{e} is close to the upper bound of \texttt{d}. With high probability, \texttt{e} does not have the minimum loss and it can also be discarded. Based on the configuration in Figure~\ref{fig:ola-loss-stop}, it is impossible to determine, exactly or approximately, which one of \texttt{b} and \texttt{d} has minimum loss. This is because \texttt{b} contains \texttt{d} and \texttt{d} is positioned almost in the center of \texttt{b}. While we have a good idea on the loss corresponding to \texttt{d}, the \texttt{b} loss is not stable yet, as reflected by the wide confidence bounds. More data have to be seen before identifying the minimum loss with high accuracy.

We design the following algorithm for stopping loss computation (Algorithm~\ref{alg:stop-loss}). The idea is to prune as many estimators as possible early in the execution. It is important to emphasize that pruning impacts only the convergence rate---not the correctness of the proposed method. Pruning conditions are exact and approximate~\cite{probTopKpruning}. All the estimators for which there exists an estimator with confidence bounds completely below their confidence bounds, can be pruned. The remaining estimators overlap. In this situation, we resort to approximate pruning. We consider three cases. First, if the overlap between the upper bound of one estimator and the lower bound of another is below a user-specified threshold, the upper estimator can be discarded with high accuracy (\texttt{a} in the example). This is a straightforward extension of the exact pruning condition. Second, if an estimator is contained inside another at the upper-end, the contained estimator can be discarded (\texttt{e} in the example). The third case is symmetric, with the inner estimator contained at the lower-end. The encompassing estimator can be discarded in this case. The algorithm is executed every time a new series of estimators are generated. Execution can be stopped when a single estimator survives the pruning process. If the process is executed until the estimators achieve the desired accuracy, we choose the estimator with the lowest expected value.

\paragraph*{When to stop gradient \& loss computation?}
Remember that gradient and loss computation are overlapped in the speculative solution we propose. When we move from a given point, we compute both the loss and the gradient at all the step sizes considered. Gradient computation is speculative since the only gradient we keep is the one corresponding to the minimum loss. The others are discarded. In the online aggregation solution, we compute estimators and confidence bounds for each of these quantities. The goal is to stop the overall computation as early as possible. How do we achieve this? We have to combine the stopping criteria for gradient and loss computation. The driving factor is loss computation. Whenever a step size can be discarded based on the exact pruning condition, the corresponding gradient estimation can be also discarded. Instead of applying the approximate pruning conditions directly, we have to consider the interaction with gradient estimation. While gradient estimation for the minimum loss does not converge, we can continue the estimation for all the step sizes that cannot be discarded based on the exact pruning condition. This allows us to identify the minimum loss and its corresponding gradient with higher accuracy.

\subsubsection{Parallel Online Aggregation}\label{sssec:est:par-ola}

In order to extend online aggregation to a parallel environment where data are partitioned across multiple processing nodes, a series of issues have to be addressed. First, how do we generate a random sample over partitioned data? The solution we use distributes data randomly across nodes at loading. It was originally proposed in~\cite{pfola:dapd}, where the authors show that a random sample is obtained by putting together samples from each of the nodes. However, notice that we do not have to gather all the samples on the same node. The processing can be executed locally on each sample and only the partial results have to merged. This is because the gradient descent problem we consider is commutative and associative. Since there is only one sample, the standard estimator applies directly. Second, how do we aggregate the partial results required for estimation? In a centralized approach, all the nodes send their local results to a designated node which merges everything together and computes the estimates. The designated node is a potential source of bottleneck when merging is time-consuming. In a distributed approach, merging is executed along an aggregation tree. While processing is faster, a delay at any of the nodes can result in delayed estimation---or even no estimation at all. And third, how and when is the online aggregation triggered? Again, two strategies are possible in a distributed environment. Online aggregation is triggered synchronously by a driver application. Or the nodes execute partial aggregation asynchronously whenever certain conditions are met, e.g., a given percentage of the local data have been processed. Notice that the actual processing across nodes is asynchronous in both situations. We argue for the synchronous approach because it provides better control over the execution: it produces the latest available estimators; it detects the stopping conditions as early as possible; and it can schedule future online aggregation requests based on the width of the confidence bounds.

\subsection{Approximate IGD}\label{ssec:ola-grad-descent:sgd}

Recall that in IGD the loss is computed only for the last $\vec{w}$ generated at the end of an iteration. This makes it impossible to detect if convergence is achieved earlier, when only a sample of the data have been used to update the model $\vec{w}$. Since faster convergence is expected in the case of massive datasets, the question we address is how to detect convergence as early as possible? This allows us to stop the current iteration immediately and start a new iteration from the latest updated model. The pseudo-code for \textit{Approximate IGD} is given in Algorithm~\ref{alg:approximate-igd}. Whenever estimator convergence is triggered, a new snapshot is taken for all the active models. Loss estimation is started for the new snapshot, executed over subsequent examples, and checked for convergence at later snapshots. A minimum number of converged loss estimators that exhibit reduced variance (Algorithm \textit{Stop IGD Loss}) is required for the process to stop.

\begin{algorithm}[htbp]
\caption{Approximate IGD}
\label{alg:approximate-igd}
\algsetup{linenodelimiter=.}

\begin{algorithmic}[1]

\FOR {\textbf{each example} $(\vec{x}_{\eta^{(j)}}, y_{\eta^{(j)}})$}

\STATE \textbf{for each \textit{active} model} $\vec{w}_{il}$ \textbf{do \textit{in parallel}}

\STATE \hspace*{0.25cm} Approximate gradient: $\nabla f \left( \vec{w}_{il}, \vec{x}_{\eta^{(j)}}; y_{\eta^{(j)}} \right)$
\STATE \hspace*{0.25cm} Update: $\vec{w}_{il} = \vec{w}_{il} - \alpha_{l} \nabla f \left( \vec{w}_{il}, \vec{x}_{\eta^{(j)}}; y_{\eta^{(j)}} \right)$

\STATE \hspace*{0.25cm} Estimate loss $L_{il}^{p}[\textit{est}, \textit{std}]$: $f(\vec{w}_{il}^{p}, \vec{x}_{\eta^{(j)}}; y_{\eta^{(j)}})$, where $p$ ranges over the snapshots

\STATE \textbf{end for}

\STATE \textbf{for each original \textit{active} model} $\vec{w}^{(k-1)}_{i}$ \textbf{do \textit{in parallel}}

\STATE \hspace*{0.25cm} Estimate loss $L_{i}[\textit{est}, \textit{std}]$: $f(\vec{w}^{(k-1)}_{i}, \vec{x}_{\eta^{(j)}}; y_{\eta^{(j)}})$

\STATE \textbf{if} test\_est\_convergence($j$) \textbf{then}

\STATE \hspace*{0.25cm} Prune out models $\vec{w}^{(k-1)}_{i}$, their children $\vec{w}_{il}$ and partial snapshots $\vec{w}_{il}^{p}$: $\textit{Stop Loss} (\{L_{i}[\textit{est}, \textit{std}]\}_{i\leq s}, .05)$

\STATE \hspace*{0.25cm} Let $t$ be the number of remaining models, i.e., \textit{active}

\STATE \hspace*{0.25cm} \textbf{if} ($t = 1$) \textbf{and} $\textit{Stop IGD Loss}$ ($\{L^{p}_{tl}[\textit{est}, \textit{std}]\}_{p<j}$, .05, 2, .01) \textbf{then} \textbf{break}

\STATE \hspace*{0.25cm} Start new snapshot $\vec{w}_{il}^{j}$ for active models 
\STATE \hspace*{0.25cm} Let $s=t$

\STATE \textbf{end if}

\ENDFOR

\end{algorithmic}

\end{algorithm}

\paragraph*{When to test for convergence?}
The two extremes to test convergence are at the end of an iteration and after every model update. The first case, i.e., standard IGD, prohibits early detection. The second case requires loss computation, i.e., a complete pass over the data, for every model update. This is impractical, unless approximation is used. We propose an intermediate solution in which convergence testing is triggered whenever certain conditions are satisfied, e.g., at fixed time intervals or when a given number of updates are processed. Convergence testing is adaptive. At the beginning of an iteration, the time interval (number of updates) is larger and it decreases as the iteration progresses, based on the loss function value. Lack of significant change in the loss between checks triggers more frequent convergence testing. This adaptive solution allows for early convergence detection, without the loss computation overhead after every update.

\paragraph*{Approximate loss computation.}
Exact loss computation at every step convergence is tested is impractical since a complete pass over the data is required. Two approximate solutions are used in practice. Notice, however, that they are used only to provide an idea on how the loss evolves, not as a convergence test for early iteration termination. In the first solution, the loss is computed only at the points where convergence is tested. The overall loss is obtained by scaling-up this loss per example value to the entire dataset size. As expected, this crude approximation is highly inaccurate and unstable. It cannot be guaranteed that progressively lower loss values are generated as more updates are processed. In the second solution, the accuracy is improved by estimating the loss over multiple data examples, situated immediately after the example where convergence is tested. The number of examples used in loss estimation is fixed. Essentially, model update and loss computation are interleaved, each operating on a distinct set of examples. There are two problems with this approach. First, there is no guarantee on the accuracy of the estimation. And second, it is not clear how to choose the number of examples used in model update and in loss computation, respectively. Intuitively, the larger the number of examples used for loss estimation, the better the accuracy. Unfortunately, this reduces the number of examples used for model update, which results in slower convergence. We propose a solution based on online aggregation to compute the loss estimate (Algorithm~\ref{alg:approximate-igd}). Given an accuracy threshold, loss estimation uses as many data examples as required in order to achieve the desired accuracy. The details are identical to loss estimation for BGD and are not shown here for conciseness. This adaptive process guarantees that an accurate estimator with provable confidence bounds is produced. Thus, convergence testing becomes theoretically sound.

\paragraph*{Overlapped model update and loss estimation.}
While online aggregation solves the accuracy issue, the number of examples derailed from model update can be too large, thus impacting convergence. Instead of the interleaved (model update-convergence testing) approach, we propose a solution in which the two stages are overlapped. Whenever convergence testing is triggered, loss estimation is started from the most recent model. The examples inspected up to the subsequent convergence testing are used to update the model and to compute the loss estimator using online aggregation. Two entities are produced at every convergence testing point---an updated model and one or more estimators for previous models. The updated model triggers the creation of a new loss estimator. If an estimator achieves the specified accuracy level, it is considered stable and is eliminated from subsequent computations. Stable estimators are used in model convergence testing. In case of convergence, the iteration is stopped and the most recent model is returned. With careful parallelization over the current multi-core CPUs, the overlapped execution strategy does not affect model update time almost at all. This is because the processing required in estimator computation is very light and the memory footprint of an estimator is in the order of a few bytes.

\begin{algorithm}[htbp]
\caption{Stop IGD Loss}
\label{alg:stop-igd-loss}
\algsetup{linenodelimiter=.}

\textbf{Input:} $\{[\textit{estimate}_{i}, \textit{std}_{i}]\}_{i\leq p}$, $\epsilon$, $m$, $\beta$

\begin{algorithmic}[1]

\STATE Let the set of converged estimators be $C =\{j \mid \frac{2\cdot \textit{std}_{j}}{\textit{estimate}_{j}} \leq \epsilon\}$

\STATE \textbf{if} ($|C| \geq m$) \textbf{and} ($\frac{\textit{max}(\textit{estimate}_{j\in C}) - \textit{min}(\textit{estimate}_{j\in C})}{\textit{max}(\textit{estimate}_{j\in C})} \leq \beta$) \textbf{then} \textbf{return} \textbf{true}
\STATE \textbf{else} \textbf{return} \textbf{false}

\end{algorithmic}

\end{algorithm}
\paragraph*{Parallel intra-iteration synchronization.}
Convergence testing in a parallel environment with data partitioned across nodes is more complicated. While all the nodes start an iteration with the same original model, as soon as the first example is considered, the models diverge. As a result, convergence testing requires partial model merging as a pre-requisite. Although this process is identical to final model merging, several questions require consideration. What do we use the merged model for? The obvious answer is loss estimation. Parallel online aggregation applies directly in this situation. The merged model can be also used as a synchronization point to re-align the local models during an iteration. This is beneficial for convergence, especially when data exhibit variation across nodes. How do we implement model synchronization? The straightforward approach is to stop model update at every node until the partial merged model is computed and passed back. The nodes are blocked over the entire duration of the distributed merging process. This is problematic because any delay is reflected in the overall execution time. The alternative is to allow the local updates to continue while the merged model is aggregated. When returned, the synchronized model is merged again with each local model to generate a new model that reflects both the global data as well as the local updates executed during merging. The merging weights are assigned proportional to the number of examples. This gives more importance to the synchronized model.

\subsection{Summary}\label{ssec:ola-grad-descent:discussion}

We introduce dynamic and adaptive intra-iteration approximation methods to estimate the gradient and the loss function. The goal is to determine when convergence is achieved as early as possible such that we can terminate the current iteration and start a new one. We expect early convergence in the case of massive training data due to redundancy. We model BGD as a series of SQL aggregate queries and apply sampling-based online aggregation to estimate gradient and loss. This allows us to compute sound estimates with theoretical confidence bounds that provide high accuracy in the termination decisions. The novelty of our approach consists in considering SQL queries with multiple aggregates -- dependent and independent -- and designing adaptive termination mechanisms that are both accurate and identify convergence early. IGD convergence detection is more difficult since the model is updated continuously over the duration of an iteration. We design a dynamic mechanism to overlap model update with loss estimation over the same example data. Convergence is accurately and timely detected since the estimation comes with theoretical guarantees. We show how our solutions are applicable to parallel environments with data partitioned across multiple processing nodes. The only requirement is a mechanism for partial aggregation.

\section{Experimental Evaluation}\label{sec:experiments}

\eat{
We evaluate the efficiency and correctness of our query overlapping and intra-iteration approximation techniques by training a Support Vector Machine (SVM) on multiple massive synthetic and real datasets. We also compare our result with fast parallel learning framework Vowpal Wabbit. Specifically, the experiments we conducted are meant to answer the following questions:

\begin{compactitem}
\item What is the effect of extensive aggregation overlapping on execution time? We show that for general machine learning tasks, e.g. SVM, stacking more aggregation together does not increase the execution time since the execution is usually bound by disk I/O.
\item What is the effect of extensive aggregation overlapping on convergence? While not imposing overhead on running time, we show that using extensive query overlapping can achieve much faster convergence.
\item What is the effect of approximation on convergence rate? With the benefit of overlapping computation, we further show that by introducing approximation, BGD and IGD are both able to even faster. It both benefit from the approximation and from detecting the convergence earlier within one iteration.
\end{compactitem}
}

The objective of the experimental evaluation is to investigate the efficiency and effectiveness of the speculative parameter testing and intra-iteration approximation techniques across several synthetic and real datasets. We consider model calibration with gradient descent optimization for two standard analytics tasks---SVM and logistic regression (LR). Moreover, we compare the efficiency of our implementation against two other distributed Big Data analytics systems---MLlib~\cite{mllib} and Vowpal Wabbit~\cite{vowpal-wabbit}. The comparison is meant to identify the overhead introduced by the proposed techniques. Specifically, the experiments we design are targeted to answer the following questions:
\begin{compactitem}
\item How does speculative parameter testing improve the convergence rate of model calibration and what overhead -- if there is any -- does it incur?
\item What is the effect of intra-iteration approximation on convergence rate and execution time? How do speculative parameter testing and intra-iteration approximation interact?
\item How do the proposed techniques stack-up against state-of-the-art Big Data analytics systems?
\item How do BGD and IGD compare when they integrate the proposed techniques?
\end{compactitem}

\subsection{Experimental Setup}\label{sec:experiments:setup}

\paragraph*{Implementation.}
We implement the speculative and online aggregation versions of distributed gradient descent optimization as GLADE PF-OLA applications. GLADE~\cite{glade:osr,glade:sigmod} is a state-of-the-art parallel data processing system that executes tasks specified using the abstract User-Defined Aggregate (UDA) interface. It was previously shown in~\cite{bismarck} that UDA is the perfect database abstraction to represent complex analytics tasks---including gradient descent optimization. Our code is thus general enough to be executed by any database supporting UDAs. To be precise, we use the code from Bismarck~\cite{bismarck} with slight modifications specific to GLADE. GLADE implements two levels of parallelism -- multi-node and multi-thread -- and takes care automatically of all the aspects related to data partitioning, task scheduling, and resource allocation. The user has to provide only the UDA code containing the model to be trained and the example data. Speculative execution is supported in GLADE through multi-query processing. Multiple instances of the same UDA -- with different parameters -- are executed against the same example data. Without going into details, we mention only that data access is shared across UDAs over the entire memory hierarchy---from disk, to memory, cache, and CPU registers. Online aggregation requires an extension of the UDA interface with estimation functions and a pre-aggregation mechanism that allows for partial aggregate computation to be triggered during query processing. These are supported by the PF-OLA framework~\cite{pfola:dapd,pfola:demo} for parallel online aggregation implemented on top of GLADE. Convergence and termination conditions are checked by the driver application. The code contains special function calls to harness detailed profiling data, used to generate the experimental results presented in the paper.

\paragraph*{System.}
We execute the experiments on a 9-node cluster running Ubuntu 12.04.4 SMP $64$-bit with Linux kernel 3.2.0-63. One node is configured as coordinator while the other eight are workers. Notice that only the workers are executing data processing tasks. Each worker has 2 AMD Opteron 6128 series 8-core processors -- 16 cores -- 28 GB of memory, and four 1 TB 7200 RPM SAS hard-drives configured RAID-0 in software. Each processor has 12 MB L3 cache while each core has 128 KB L1 and 512 KB L2 local caches. The storage system supports 240, 420 and 1600 MB/second minimum, average, and maximum read rates, respectively---based on the Ubuntu disk utility. There are two file systems running on each node. The storage system containing experimental data is mounted as a local file system, while the code and executables are shared across nodes through an NFS file system instance.

\paragraph*{Methodology.}
We perform all experiments at least 3 times and report the average value as the result. We always enforce data to be read from disk in the first iteration by cleaning the file system buffers before execution. Subsequent iterations can access cached data. When the execution time per iteration is reported, the value corresponds to iterations two and above. We execute each algorithm for a fixed number of iterations---typically 20.

\begin{table}[htbp]
  \begin{center}
    \begin{tabular}{l||r|r|r}

	\textbf{Dataset} & \textbf{Dimensions} & \textbf{Examples} & \textbf{Size}\\

	\hline
	
	\texttt{forest}~\cite{bismarck} & 54 & 581K & 485 MB\\
	
	\texttt{classify50M}~\cite{bismarck} & 200 & 50M & 136 GB\\

	\texttt{splice}~\cite{vowpal-wabbit} & 13M & 50M & 3.2 TB\\

    \end{tabular}
  \end{center}

\caption{Datasets used in the experiments.}\label{tbl:datasets}
\end{table}

\paragraph*{Tasks and datasets.}
While gradient descent is a general optimization method that can be applied for training a large variety of models, in this paper, we present experiments for convex LR and SVM. We use three datasets -- two real and one synthetic, i.e., \texttt{classify50M} -- with very different characteristics, as depicted in Table~\ref{tbl:datasets}. It is important to emphasize that, as far as we know, \texttt{splice} is the largest dataset available on which gradient descent results have been published in the literature---in machine learning and databases. Since \texttt{forest} has small size, we use it to evaluate the performance of speculative processing on a single machine. Only multi-threading parallelism is used in this case. The other two datasets are evenly partitioned across the 8 worker nodes in the cluster using a random hash function. BGD and multiple versions of IGD are implemented for each (model type, dataset) combination. \texttt{IGD merge} corresponds to model averaging, in which a separate model is created for every thread in the system. \texttt{IGD lock} uses a single model per node and synchronizes access across threads with a light locking mechanism, e.g., \texttt{compare\&swap}. \texttt{IGD no lock} implements the solution proposed in~\cite{nolock-igd} that eliminates synchronization. While we run experiments for all the combinations, we include in the paper only the most representative results.

\begin{figure*}[htbp]
\begin{center}
\subfloat[]{\includegraphics[width=0.33\textwidth]{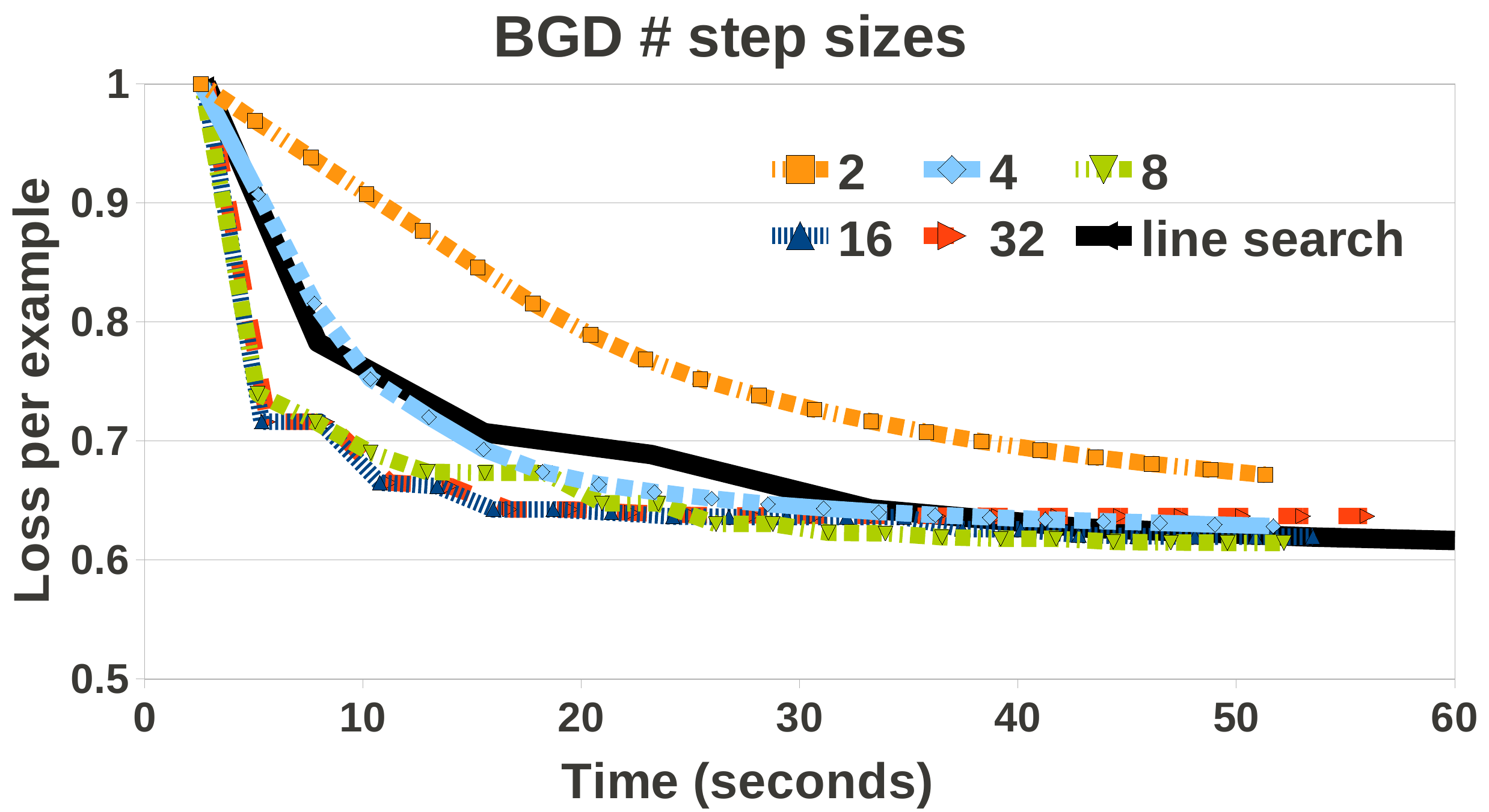}\label{fig:forest:bgd:svm}}
\subfloat[]{\includegraphics[width=0.33\textwidth]{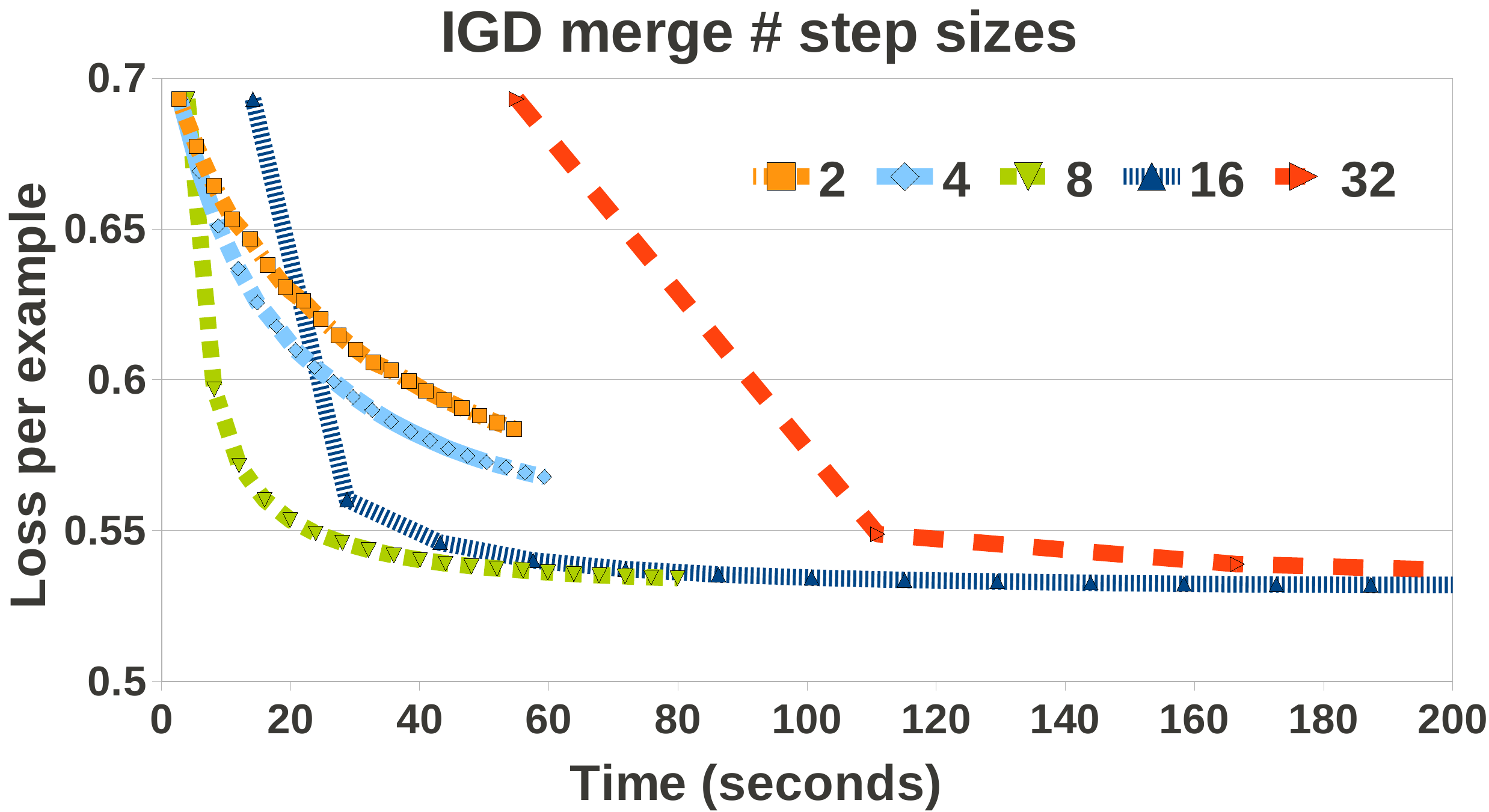}\label{fig:forest:igd:lr}}
\subfloat[]{\includegraphics[width=0.33\textwidth]{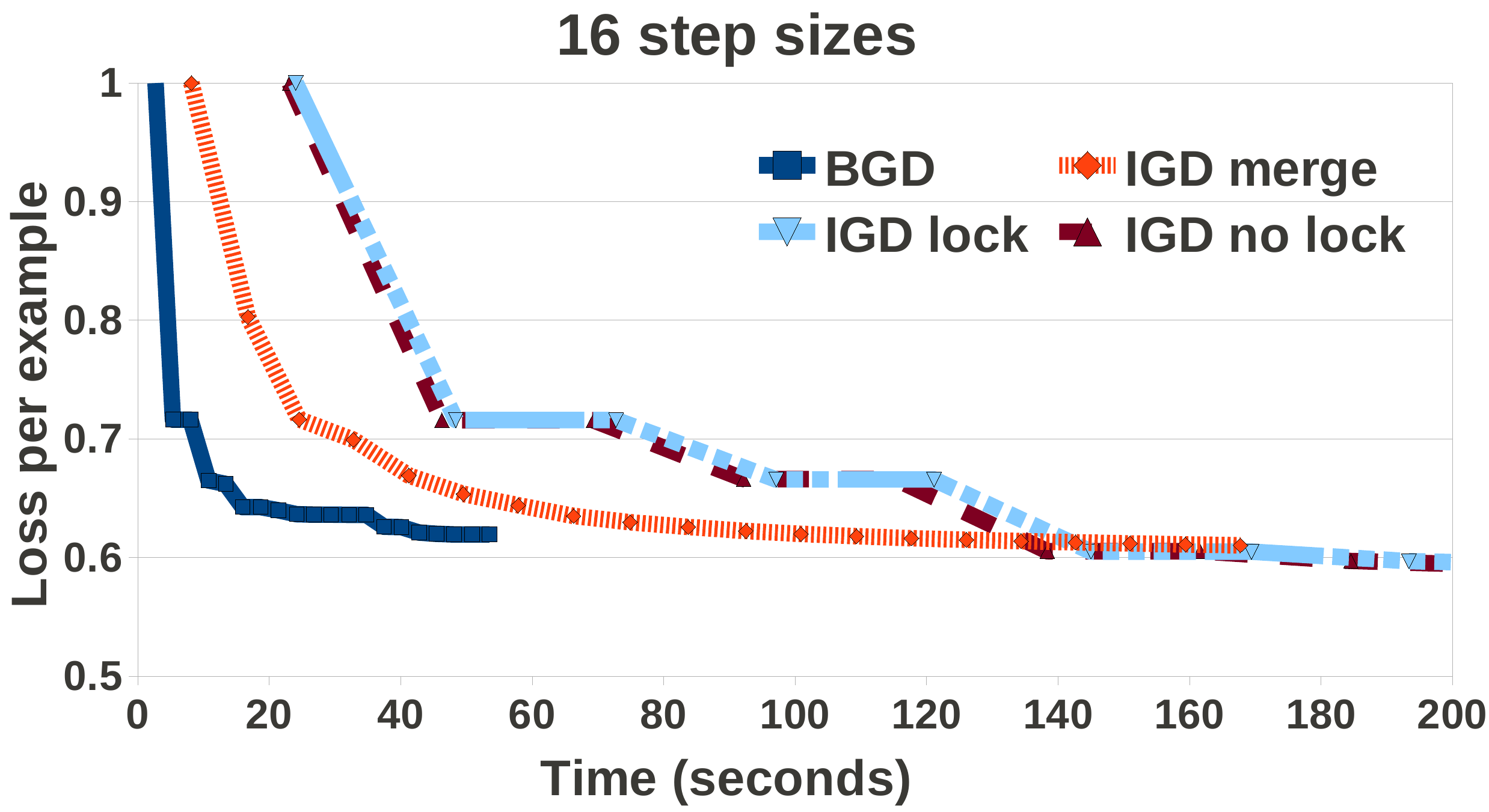}\label{fig:forest:convergence:svm}}
\caption{Convergence as a function of the number of step sizes on \texttt{forest}. (a) SVM with BGD. (b) LR with IGD. (c) SVM.}
\label{fig:forest:convergence}
\end{center}
\end{figure*}

\subsection{Speculative Parameter Testing}\label{sec:experiments:speculative}

\paragraph*{Convergence rate.}
To quantify the impact of speculative parameter testing on convergence rate, we execute BGD and IGD with an increasing number of concurrent step sizes. We pick the step size values by mimicking a real parameter tuning procedure. We start with an arbitrary value and then add smaller and larger values. The old values are maintained when increasing the number of steps. This guarantees continuous improvement as the number of steps increases---not always the case for Bayesian inference. Notice though that Bayesian inference can provide better values that result in faster convergence. Figure~\ref{fig:forest:convergence} depicts the effect of increasing the number of step sizes on convergence for the \texttt{forest} dataset. The general trend is to achieve faster convergence as the number of step sizes increases. While this is always true for BGD, the IGD behavior is more nuanced. The BGD results are depicted in Figure~\ref{fig:forest:bgd:svm}. They include a comparison with \texttt{line search}~\cite{convex-optimization}, the standard backtracking step size search method that guarantees a certain loss decrease rate at every iteration. \texttt{line search} adjusts the step size at every iteration, thus providing automatic step size tuning. For more than 4 step sizes, speculative processing achieves faster convergence than \texttt{line search}. This is because \texttt{line search} limits itself to satisfying the imposed loss decrease rate. For a higher decrease rate, \texttt{line search} requires more passes over the example data, which results in higher time per iteration, thus slower convergence. Figure~\ref{fig:forest:igd:lr} depicts the convergence rate for \texttt{IGD merge}. While a higher number of step sizes still generates a higher loss decrease per iteration, the overhead incurred is significantly higher in this case. By the time one iteration with 32 steps finishes, 8 steps has already finished execution and achieved convergence. A direct comparison between BGD and the IGD versions for 16 step sizes is depicted in Figure~\ref{fig:forest:convergence:svm}. It is clear that BGD can handle a large number of step sizes better than IGD since the number of speculative models is an order of complexity lower, i.e., linear vs. quadratic. Between the IGD solutions, \texttt{IGD merge} clearly outperforms the other two versions, even though the convergence rate per iteration is higher for \texttt{IGD lock} and \texttt{IGD no lock}, respectively. The hardware cache coherence mechanism coupled with the time to update a large number of models are the reason for the poor behavior of \texttt{IGD no lock}.

\begin{table*}[htbp]
  \begin{center}
    \begin{tabular}{l|l||rrrrr|rrrrr|rr}

	\multicolumn{2}{c||}{} & \multicolumn{5}{|c|}{\texttt{forest}} & \multicolumn{5}{c|}{\texttt{classify50M}} & \multicolumn{2}{c}{\texttt{splice}} \\
	
	\multicolumn{2}{c||}{} & 1 & 2 & 4 & 16 & 32 & 1 & 2 & 4 & 16 & 32 & 1 & 2 \\

	\cline{3-14}

	\multirow{4}{*}{SVM} & BGD & 2.57 & 2.58 & 2.57 & 2.71 & 2.83 & 95 & 95 & 96 & 97 & 99 & 625 & 639 \\
	
	& IGD merge & 2.58 & 2.58 & 2.74 & 8.17 & 33.2 & 96 & 96 & 99 & 363 & 1749 & 631 & 675 \\

	& IGD lock & 2.8 & 2.91 & 3.12 & 24.04 & 92.78 & 95 & 96 & 101 & 880 & 3190 & 840 & 1064 \\

	& IGD no lock & 2.5 & 2.52 & 3.06 & 23.12 & 86.73 & 95 & 96 & 101 & 726 & 2834 & 631 & 703 \\

	\hline

	\multirow{4}{*}{LR} & BGD & 2.65 & 2.65 & 2.66 & 2.92 & 3.09 & 96 & 96 & 96 & 97 & 99 & 618 & 640 \\
	
	& IGD merge & 2.67 & 2.67 & 2.85 & 14.15 & 55.02 & 96 & 96 & 99 & 432 & 2213 & 634 & 1039 \\

	& IGD lock & 2.8 & 2.8 & 3.1 & 27.22 & 104.24 & 97 & 97 & 102 & 898 & 3301 & 2103 & 2199 \\

	& IGD no lock & 2.71 & 2.71 & 3.06 & 26.69 & 101.04 & 97 & 97 & 101 & 772 & 2975 & 636 & 1023 \\

    \end{tabular}
  \end{center}

\caption{Execution time per iteration for multiple step sizes across all the datasets and all the methods considered in the experiments.}\label{tbl:datasetsTime}
\end{table*}

\paragraph*{Overhead.}
Table~\ref{tbl:datasetsTime} contains the execution time per iteration for all the experimental configurations. In the case of the \texttt{splice} dataset, for more than two step sizes, the memory required by the model is beyond the physical capacity of the testing machine. The time per iteration changes minimally for speculative BGD when we increase the number of step sizes from 1 to 32. The slightly higher execution time for LR is due to the more complicated gradient computation. IGD incurs a considerable overhead when the number of step sizes increases since the computation is quadratic in the number of step sizes. Between the IGD versions, \texttt{IGD merge} is the most efficient, especially for a large number of step sizes. This is due to complete model replication across threads which eliminates contention. Overall, speculative parameter testing is able to boost the convergence rate for BGD up to 32 step sizes, without significantly increasing the execution time per iteration. We are able to achieve this because our implementation takes full advantage of the parallelism available in modern multi-core CPUs, including deep pipelines and vectorized instructions. The main idea is to execute all the processing -- across all the models -- with minimal data movement, i.e., whenever a data example is brought in the CPU registers, it is used to update all the gradients/models.

\begin{figure*}[htbp]
\begin{center}
\subfloat[]{\includegraphics[width=0.33\textwidth]{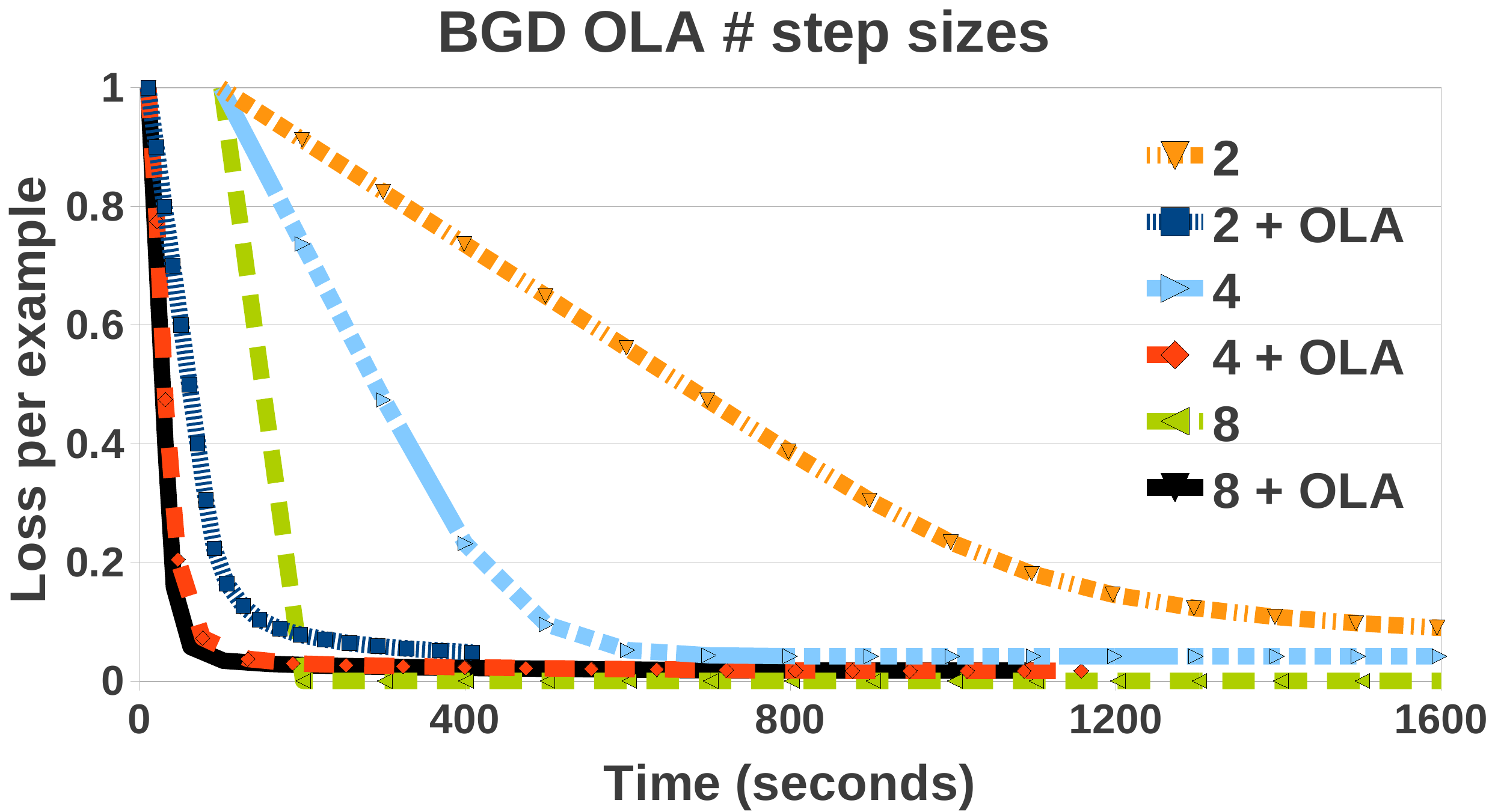}\label{fig:ola:svm:classify}}
\subfloat[]{\includegraphics[width=0.33\textwidth]{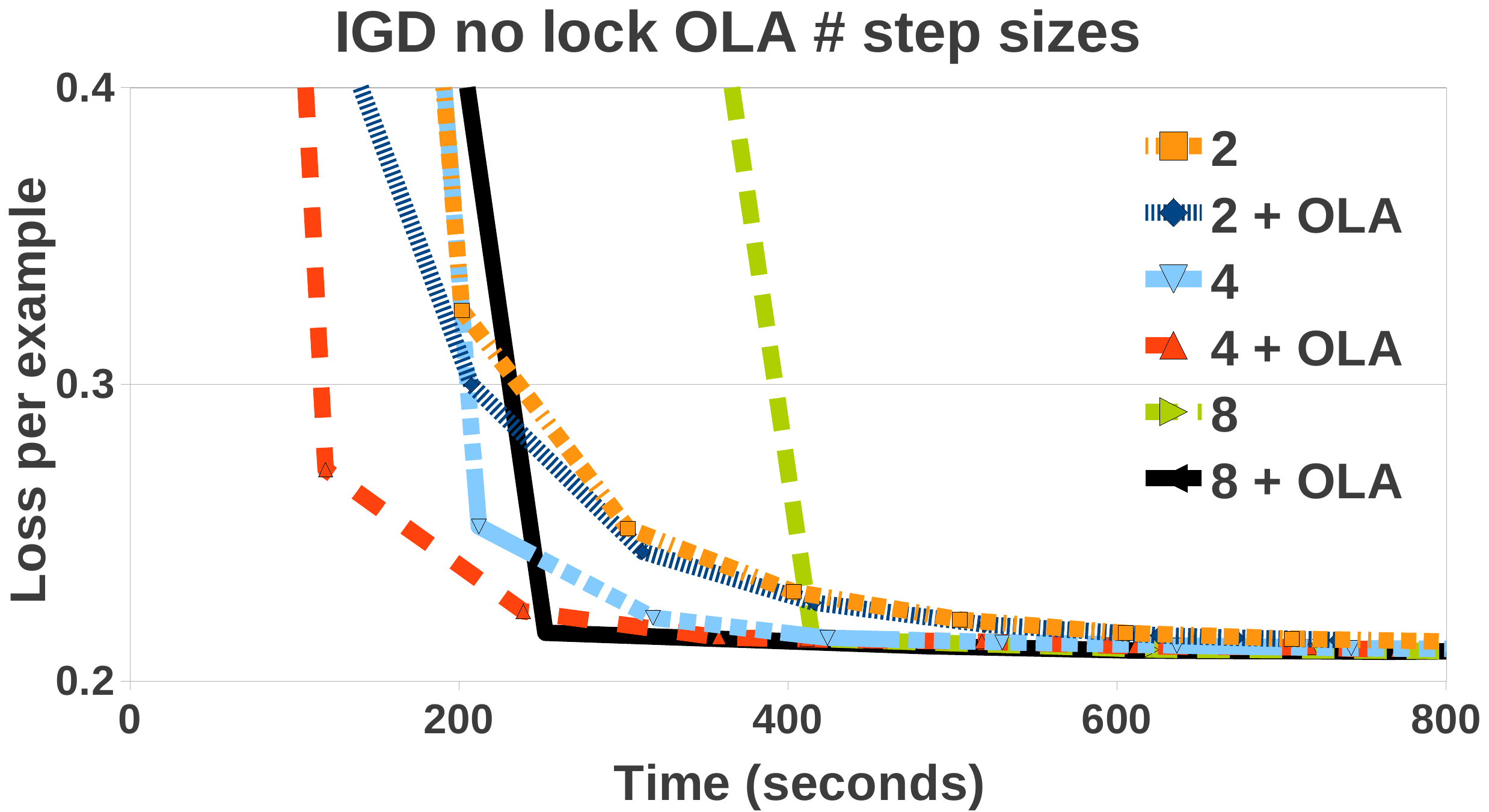}\label{fig:ola:lr:classify}}
\subfloat[]{\includegraphics[width=0.33\textwidth]{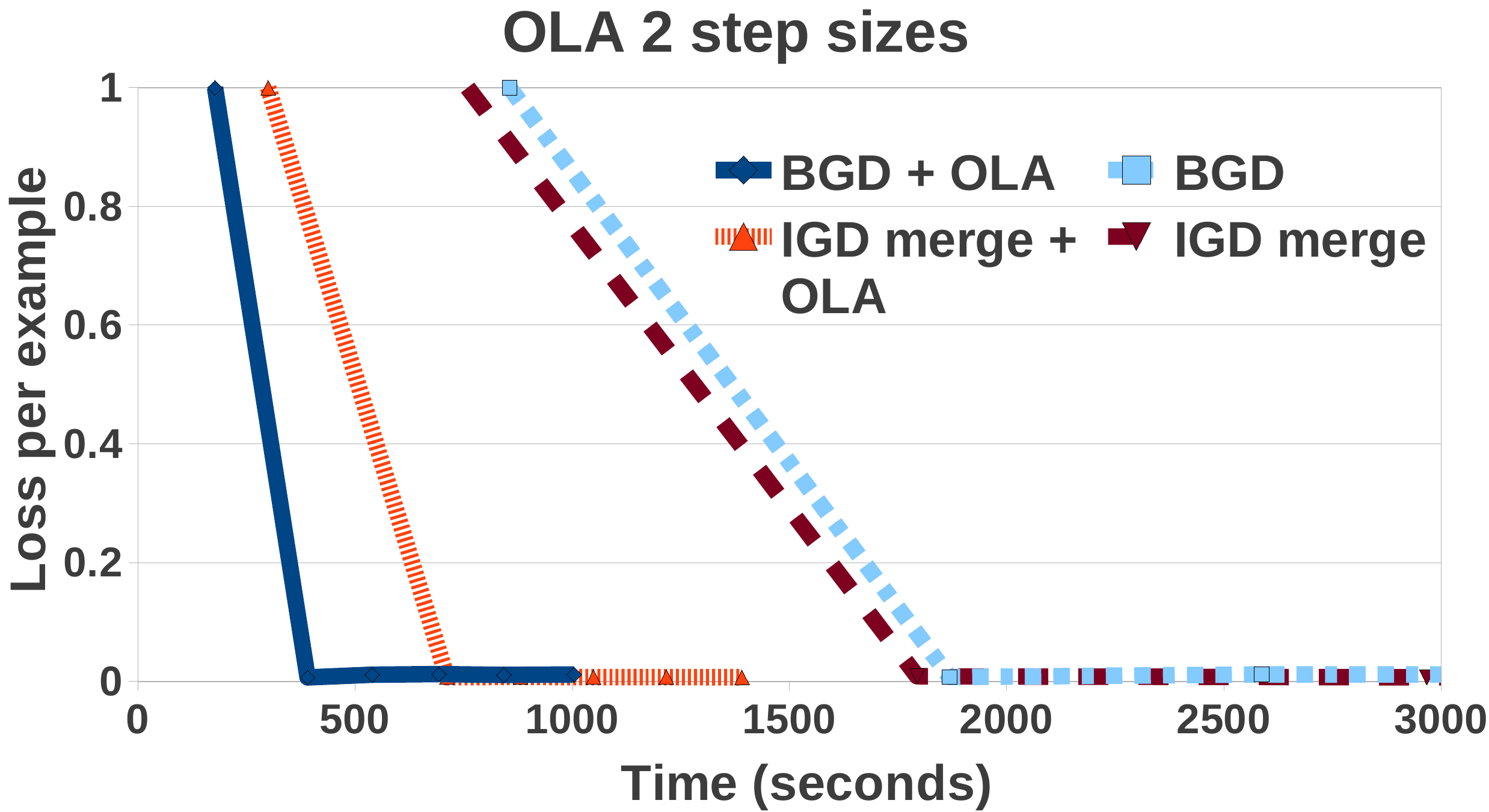}\label{fig:ola:svm:splice}}
\caption{Convergence as a function of online aggregation on \texttt{classify50M}. (a) SVM with BGD. (b) LR with IGD. (c) SVM (\texttt{splice}).}
\label{fig:ola:convergence}
\end{center}
\end{figure*}

\subsection{Online Aggregation}\label{sec:experiments:ola}

The combined effect of online aggregation and speculative parameter testing on convergence rate is depicted in Figure~\ref{fig:ola:convergence}. The execution of an iteration is halted as soon as the width of the confidence bounds corresponding to the estimator is below 5\% of the estimate, i.e., we are 95\% confident on the estimator value. In the case of BGD (Figure~\ref{fig:ola:svm:classify}), online aggregation provides a considerable boost in convergence rate across all the tested step sizes. The gradient can be estimated accurately from a small sample. The same is true for the loss. This allows for immediate detection of promising step sizes, while the sub-optimal ones can be discarded. In the best case, convergence to the same or a better loss is achieved as much as 20 times faster. The same trend can be observed for IGD (Figure~\ref{fig:ola:lr:classify}). The benefit of online aggregation is the most obvious for 8 step sizes, since convergence is achieved within one iteration. With standard IGD, convergence can be detected only at the end of a complete pass over the training data. Due to partial model merging, online aggregation detects convergence much earlier. Figure~\ref{fig:ola:svm:splice} depicts a direct comparison between BGD and IGD with online aggregation enabled for the \texttt{splice} dataset. While IGD achieves slightly faster convergence in the standard case, BGD outperforms IGD by more than 50\% when online aggregation is enabled. This is because gradient estimation is a considerably easier task than detecting convergence for partial models.

\begin{figure}[htbp]
\begin{center}
{\includegraphics[width=0.75\textwidth]{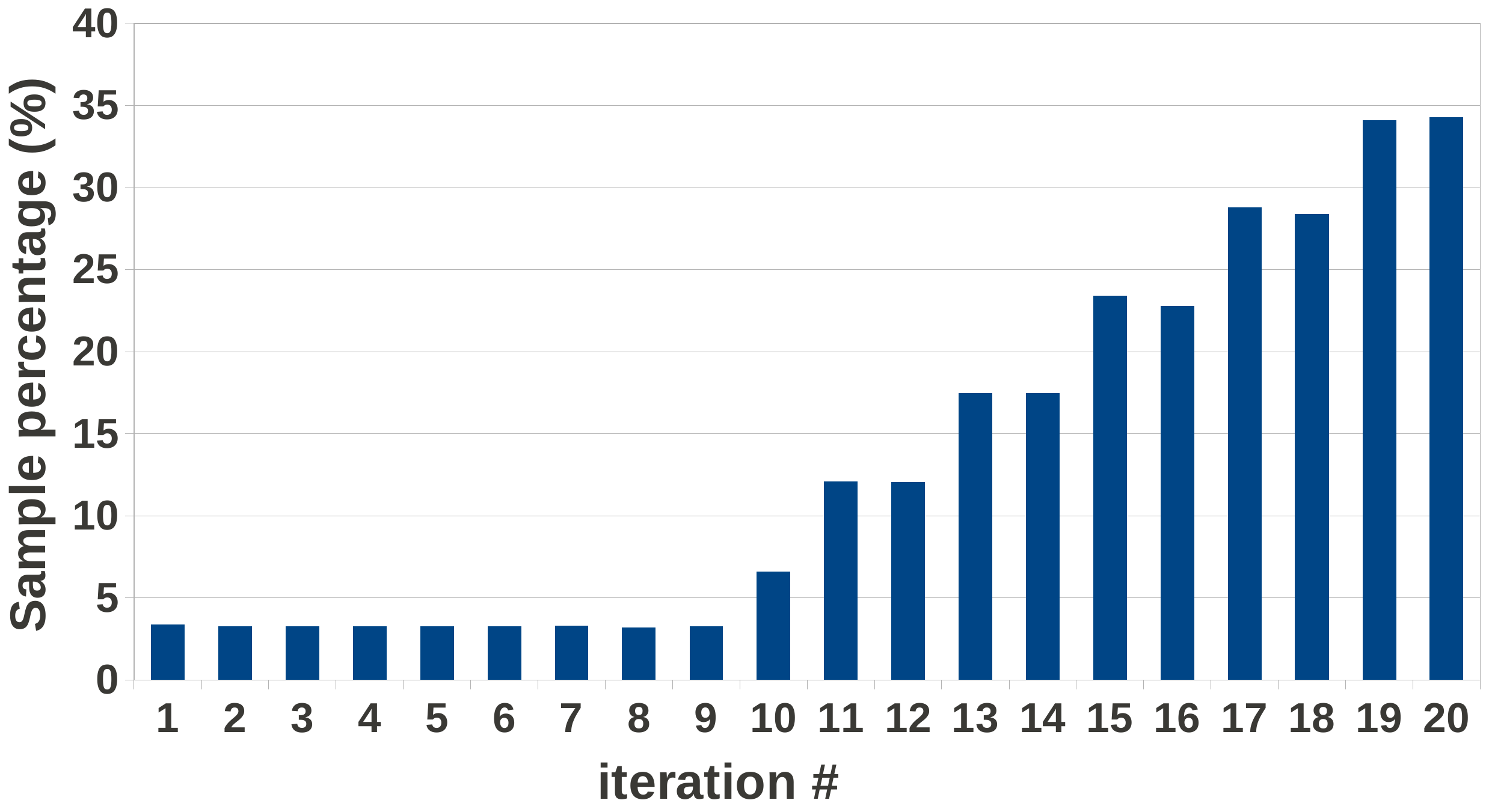}}
\caption{Percentage of samples per iteration.}
\label{fig:sample}
\end{center}
\end{figure}

\paragraph*{Adaptive sampling.}
In order to confirm that online aggregation is the appropriate solution for intra-iteration approximation -- and not sub-sampling with a pre-determined size -- we measure the sample size required for the estimators to achieve convergence in each iteration. Figure~\ref{fig:sample} depicts the sampling ratio when an iteration is halted. While the sampling ratio is well below 5\% in the first iterations, it increases drastically once we are getting close to the minimum. This happens because more data have to be seen for the estimators to converge. If a fixed-size sub-sample is taken and model calibration is executed on the sub-sample, two events can occur. If the sample size is too large, unnecessary processing is executed. For a smaller sample, convergence to the true minimum cannot be achieved, no matter how long the process is executed. Since online aggregation chooses the size of the sample adaptively, it avoids these problems altogether.

\begin{figure*}[htbp]
\begin{center}
\subfloat[]{\includegraphics[width=0.45\textwidth]{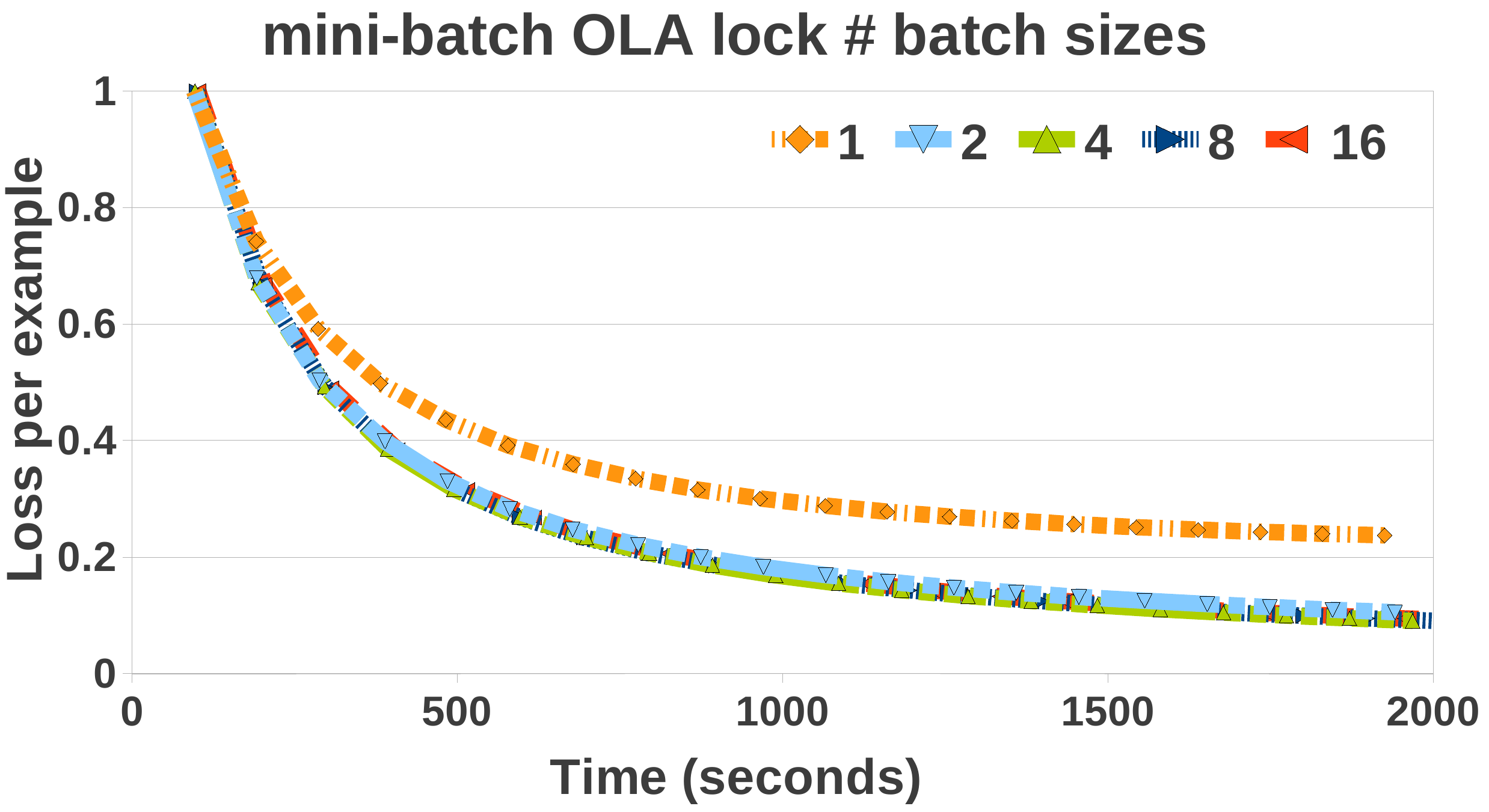}\label{fig:lr:classify:batchsizes}}
\subfloat[]{\includegraphics[width=0.45\textwidth]{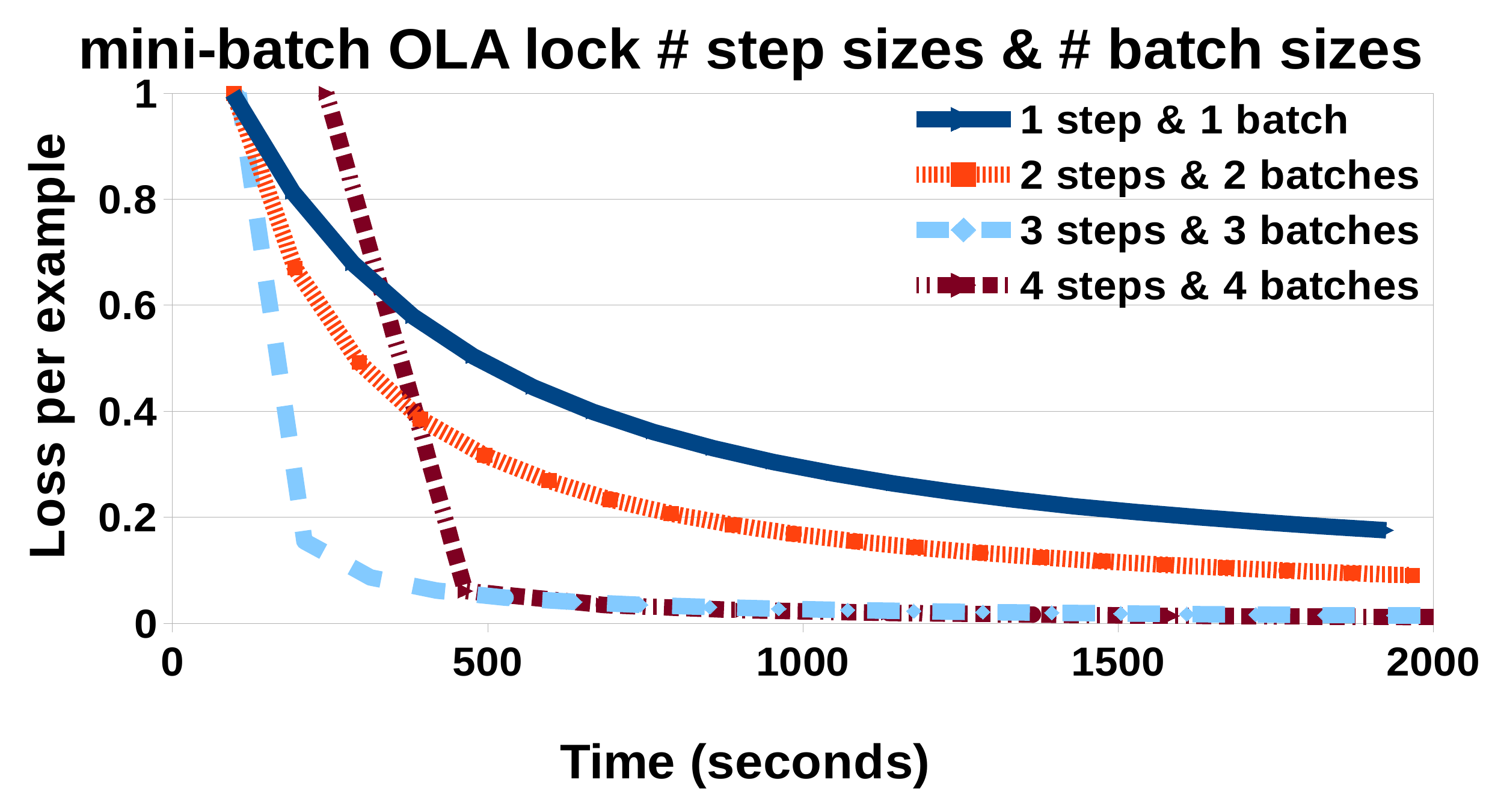}\label{fig:lr:classify:stepsize-batchsize}}
\caption{Convergence as a function of batch and step size for mini-batch OLA on \texttt{classify50M}. (a) LR with fixed step size and different batch sizes. (b) LR with different step sizes and different batch sizes.}
\label{fig:stepsize-batchsize}
\end{center}
\end{figure*}

\subsection{Two-Parameter Experiments}\label{sec:experiments-multi}

We examine how the proposed techniques apply to a scenario where two parameters have to be calibrated. Specifically, we consider mini-batch gradient descent (Section~\ref{ssec:grad-descent:stochastic}) as a variant of IGD in which the gradient is approximated using a variable number of examples, rather than a single example. The number of examples, i.e., the batch size, is the second parameter -- together with the step size -- to tune. Since these two parameters are correlated -- the larger the batch size, the larger the step size -- we draw their values from a single 2-D normal distribution characterized by a covariance matrix that reflects this relationship. The normal distribution is initialized with centers at 0.1 (step size) and 1000 (batch size), variances of 0.1 and 10000, respectively, and positive covariance of 10. The distribution is updated after every iteration using Bayesian inference.

Figure~\ref{fig:stepsize-batchsize} depicts the convergence behavior of mini-batch gradient descent as a function of the two parameters considered---step size and batch size. In Figure~\ref{fig:lr:classify:batchsizes}, we use a single constant step size, while we vary the number of batch sizes. This allows us to isolate the batch size effect. We observe that there is little difference beyond two batch sizes. Figure~\ref{fig:lr:classify:stepsize-batchsize} illustrates the combined effect of the two parameters. As expected, a larger number of configurations results in faster convergence rate---limited only by the higher update time corresponding to \texttt{IGD lock}. When comparing the two figures, we conclude that the step size is the main factor impacting convergence. This confirms our approach focused on calibrating the step size. It is worth noting that step size and batch size are the most important tunable parameters for gradient descent methods applied to any type of objective functions. While our methods are directly applicable to a larger number of parameters, it is difficult to identify the effect each parameter has on convergence when bundled together with other parameters.

\subsection{Comparison with Existing Systems}\label{sec:experiments:comp}

We compare our implementation with two state-of-the-art large scale analytics systems -- MLlib~\cite{mllib} and Vowpal Wabbit~\cite{vowpal-wabbit} -- to validate the efficiency of our solutions. We choose these systems based on the following criteria. They are disk-based distributed systems with native support for gradient descent optimization. In fact, they support only IGD and not BGD. Based on the complete study by Cai et al.~\cite{jermaine:largeML:comp}, no other open-source system satisfies these conditions. Our goal is to prove that the improvement we get from speculative processing and intra-iteration approximation does not come from an inefficient implementation. In fact, as shown in Table~\ref{tbl:comp}, our implementation is faster than these systems. The measure we use is the time per iteration for executing a complete gradient update pass and a complete loss computation.

\begin{table}[htbp]
  \begin{center}
    \begin{tabular}{l|l||r|r|r}

	\multicolumn{2}{c||}{} & \multicolumn{1}{|c|}{\texttt{VW}} & \multicolumn{1}{c|}{\texttt{MLlib}} & \multicolumn{1}{c}{\texttt{GLADE PF-OLA}} \\
	

	\cline{3-5}

	\multirow{2}{*}{\texttt{classify50M}} & SVM & 248 & 180 & 96  \\
	
	& LR & 446 & 180 & 96  \\

	\hline

	\multirow{2}{*}{\texttt{splice}} & SVM & 1256 & 70560 & 631  \\
	
	& LR & 3100 & 70560 & 634  \\

    \end{tabular}
  \end{center}

\caption{Execution time per iteration (seconds).}\label{tbl:comp}
\end{table}

\paragraph*{MLlib.}
We run MLlib over Spark on the same 9-node cluster. We set 4 workers per node and 4 cores per worker, which takes-up all the 16 cores in a node. We assign 6 GB of memory per worker, i.e., 24 GB out of the total 28 GB of memory per node. In all the MLlib results, we do not consider the data loading time for Spark, which is considerably larger than the time reported in Table~\ref{tbl:comp}. Nonetheless, we do include the data reading time from disk in our solution. For \texttt{classify50M}, MLlib is capable to cache the entire dataset in memory, thus it has a decent execution time. This is not the case for \texttt{splice} and the results show it.

\paragraph*{Vowpal Wabbit (VW).}
VW is I/O-bound if the input data are pre-loaded in its native binary format. The results reported in Table~\ref{tbl:comp} are for loaded data. VW requires an additional pass over the data to compute the loss exactly. The estimated loss it provides at each node during an iteration is local. The time for a single pass over the data is half of the results shown in Table~\ref{tbl:comp}. While closer to the results obtained by our implementation, it is important to remember that we overlap model update and loss computation.

\paragraph*{GLADE PF-OLA.} Notice that we measure the execution time of the naive IGD implementation which does not include speculative processing and intra-iteration approximation. If we consider speculative processing, i.e., multiple concurrent step sizes, the execution time for GLADE PF-OLA stays almost the same (Table~\ref{tbl:datasets}), while MLlib and VW require $n$ times longer to complete, where $n$ is the number of steps. Likewise, if we consider intra-iteration approximation, the execution time per iteration is even shorter, as shown in Figure~\ref{fig:ola:convergence}.

\subsection{Discussion}\label{sec:experiments:discussion}

We validate the effectiveness and efficiency of the proposed techniques -- speculative processing and intra-iteration approximation -- for large scale gradient descent optimization. We show that speculative processing speeds-up the convergence rate of both BGD and IGD, and online aggregation reduces the iteration time further. We find that BGD is better suited to integrate the proposed techniques. In our experiments, BGD always outperforms IGD for similar settings. By comparing with state-of-the-art systems, we confirm that our solution is able to significantly boost the model calibration time for terascale datasets.

\section{Related Work}\label{sec:rel-work}

We discuss related work in two main categories: distributed gradient descent optimization and parallel online aggregation. We emphasize the novelty brought by this work when compared to our previous papers on parallel online aggregation~\cite{pfola:dapd,pfola:demo} and incremental gradient descent in GLADE~\cite{igd-glade,lmf-hybrid-glade}.

\paragraph*{Distributed gradient descent optimization.}
There is a plethora of work on distributed gradient descent algorithms published in machine learning~\cite{distributed-mini-batch,google-brain,parallel-igd,distributed-lmf}. All these algorithms are similar in that a certain amount of model updates are performed at each node, followed by the transfer of the partial models across nodes. The differences lie in how the model communication is done~\cite{distributed-mini-batch,parallel-igd} and how the model is partitioned for specific tasks, e.g., neural networks~\cite{google-brain} and matrix factorization~\cite{distributed-lmf}. Many of the distributed solutions are immediately applicable to multi-core shared memory environments. The work of R{\'e} et al.~\cite{nolock-igd,bismarck,dimmwitted} is representative in this sense. Our work is different from all these approaches because we consider concurrent evaluation of multiple step sizes and we use adaptive intra-iteration approximation to detect convergence. Moreover, IGD is taken by default to be the optimal gradient descent method, while BGD is hardly ever considered. We provide a thorough comparison between BGD and IGD, and show that -- with our optimizations -- BGD always outperforms IGD.

Many Big Data analytics systems and frameworks implement gradient descent optimization. Most of them target distributed applications on top of the Hadoop MapReduce framework, e.g., Mahout~\cite{mahout}, MLlib~\cite{mllib}, while others provide complete stacks, e.g., MADlib~\cite{madlib}, Distributed GraphLab~\cite{graphlab-distributed}, and Vowpal Wabbit~\cite{vowpal-wabbit}. With no exception, IGD is the only method implemented in all these systems. As a first step, the techniques we present in this paper can be incorporated into any of these systems, as long as multi-threading parallelism and partial aggregation are supported. More important, we provide strong evidence that BGD deserves full consideration in any Big Data analytics system.

\paragraph*{Online aggregation.}
The database online aggregation literature has its origins in the seminal paper by Hellerstein et al.~\cite{ola}. We can broadly categorize this body of work into system design~\cite{control,dbo,demo:dbo,turbo:dbo}, online join algorithms~\cite{ripple-join,sms-join,pr-join}, online algorithms for estimations other than join~\cite{ola-set,ola-extreme}, and methods to derive confidence bounds~\cite{haas-bounds}. All of this work is targeted at single-node centralized environments. The parallel online aggregation literature is not as rich though. We identified only three lines of research that are closely related to this paper. Luo et al.~\cite{par-hash-ripple} extend the centralized ripple join algorithm~\cite{ripple-join} to a parallel setting. A stratified sampling estimator~\cite{sampling-techniques} is defined to compute the result estimate while confidence bounds cannot always be derived. Wu et al.~\cite{distributed-ola} extend online aggregation to distributed P2P networks. They introduce a synchronized sampling estimator over partitioned data that requires data movement from storage nodes to processing nodes. In subsequent work, Wu et al.~\cite{continuous-sampling} tackle online aggregation over multiple queries. The third piece of relevant work is online aggregation in MapReduce. In~\cite{hadoop-online}, standard Hadoop is extended with a mechanism to compute partial aggregates. In subsequent work~\cite{MR-ola}, an estimation framework based on Bayesian statistics is proposed. BlinkDB~\cite{blink,agarwal:bootstrap} implements a multi-stage approximation mechanism based on pre-computed sampling synopses of multiple sizes, while EARL~\cite{earl} and ABS~\cite{zeng:bootstrap} use bootstrapping to produce multiple estimators from the same sample.

\paragraph*{GLADE PF-OLA.}
The novelty of our work compared to the PF-OLA framework~\cite{pfola:dapd,pfola:demo} comes from applying online aggregation estimators to complex analytics, rather than focusing on standard SQL aggregates---the case in previous literature. We are the first to model gradient descent optimization as an aggregation problem. This allows us to design multiple concurrent estimators and to define halting mechanisms that stop the execution when model update and loss computation are overlapped. Moreover, the integration of online aggregation with speculative step evaluation allows for early identification of sub-optimal step sizes and directs the system resources toward the promising configurations. None of the existing systems, including GLADE PF-OLA, support concurrent hyper-parameter evaluation or concurrent estimators. Our previous work on gradient descent optimization in GLADE~\cite{igd-glade,lmf-hybrid-glade} is limited to IGD. In this paper, we also consider BGD and propose general methods applicable to distributed gradient descent optimization.

\section{Conclusions and Future Work}\label{sec:conclusions}

In this paper, we propose two database techniques for efficient model calibration. Speculative parameter testing allows for several parameter configurations to be evaluated concurrently. Online aggregation identifies sub-optimal configurations early in the processing. We apply the proposed techniques to distributed gradient descent optimization -- batch and incremental -- and provide an extensive experimental comparison between these two methods. Contrary to the general belief, BGD always outperforms IGD both in convergence speed and execution time. In future work, we plan to extend the proposed techniques to other model calibration methods beyond gradient descent.

\paragraph*{Acknowledgments.}
The work in this paper is supported by a Hellman Faculty Fellowship and a gift from LogicBlox.

\bibliographystyle{abbrv}

\end{document}